\documentclass[letterpaper, english,reprint,prl]{revtex4-2}
\usepackage[T1]{fontenc}
\usepackage[latin9]{inputenc}
\usepackage{xcolor}
\usepackage{textcomp}
\usepackage{amsmath}
\usepackage{amssymb}
\usepackage{tikz}
\usepackage{natbib}
\usepackage{mhchem} 

\makeatletter

\pdfpageheight\paperheight
\pdfpagewidth\paperwidth

\usepackage{tikz}

\makeatother

\usepackage{babel}
\begin{document}
\preprint{APS/123-QED}
\title{Microscopic Theory of Nonlinear Hall Effect in Three-dimensional Magnetic
Systems}
\author{Wen-Tao Hou}
\email{houwentao@tiangong.edu.cn}

\affiliation{School of Physical Science and Technology, Tiangong University, Tianjin,300387,
China}

\author{Jiadong Zang}
\email{jiadong.zang@unh.edu}

\affiliation{Department of Physics and Astronomy, University of New Hampshire,
Durham, New Hampshire 03824, USA}
\begin{abstract}
The nonlinear Hall effect (NLHE) has been detected in various of condensed
matter systems. Unlike linear Hall effect, NLHE may exist in physical
systems with broken inversion symmetry in the crystal. On the other
hand, real space spin texture may also break inversion symmetry and
result in NLHE. In this letter, we employ the Feynman diagrammatic
technique to calculate nonlinear Hall conductivity (NLHC) in three-dimensional
magnetic systems. The results connect NLHE with the physical quantity
of emergent electrodynamics which originates from the magnetic texture.
The leading order contribution of NLHC $\chi_{abb}$ is proportional
to the emergent toroidal moment $\mathcal{T}_{a}^{e}$ which reflects
how the spin textures wind in three dimension. 

\begin{description}
\item [{PACS~numbers}] 72.80.-r, 75.10.\textminus b, 75.47.-m, 75.76.+j 
\item [{Keywords}] Nonlinear Hall conductivity, emergent gauge field, magnetic
Hopfion
\end{description}
\end{abstract}
\maketitle
\begin{figure}
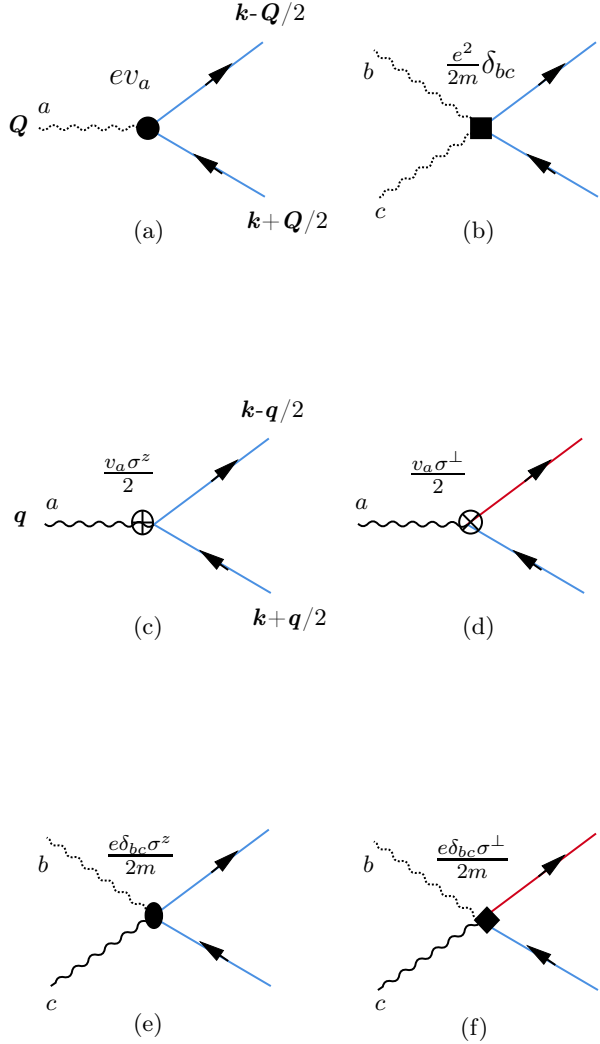

\include{FeynRule}\caption{Feynman Rules\label{fig:Feynman-Rules}.}
\end{figure}
\begin{figure}
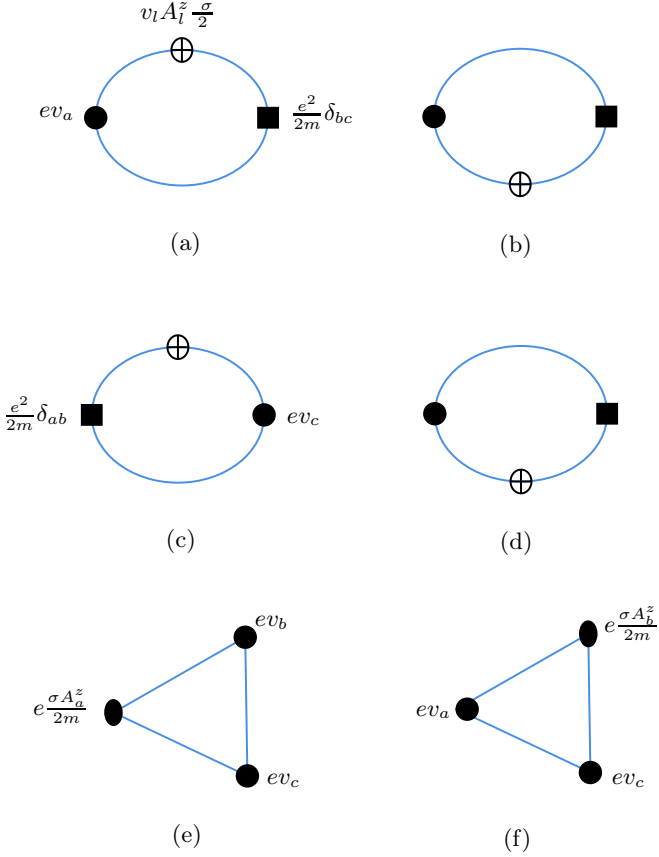

\include{no_switch}

\caption{Nonzero leading order diagrams without spin flip\label{fig:no-switch}}
\end{figure}

Plethora of Hall effects have unveiled many exciting physical phenomena
in condensed matter physics, such as the quantum spin hall effect,
which reveals the non-trivial topological band structures of topological
insulators\citep{QSHE2006}. Unlike linear Hall effects, the non-linear
Hall effect (NLHE) would happen in the systems without time reversal
symmetry broken\citep{NonlinearTRS,NLHTRS}. Its close connection to
non-trivial topology makes NLHE as an useful tool to unlock novel
condensed matter systems, such as Weyl semimetals\citep{NLHWeylTheory2021,Machon2022Weyl}
and topological insulators\citep{Yasuda2017Ti,Rao2021Ti}. Nonlinear
Hall conductivty (NLHC) $\chi_{abb}$ has been measured in layered
WTe$_{2}$ structures\citep{QiongMaWTe22019}, layered graphene\citep{NLHGraphene2,NLHGraphene2022},
Weyl semimetal TaIrTe$_{4}$\citep{WeylTaIrTe42021} and topological
antiferromagnetic heterostructure\citep{NHAnti2023}. Theoretically,
several technical methods have been developed to compute $\chi_{abb}$
in real materials, such as Boltzmann transport theory\citep{NLHBolz2019,NLHBoltz2019}, 
Feynman diagrammatic technique\citep{NLHDiagramMoore2019,NLHGeneralHaizhou2021,NLHGeneral2}
and numerical supercell method\citep{Chen2024Lattice}. It
has been revealed that the NLHC is associated with the Berry curvature
dipole in the reciprocal space \citep{Gao2014Field,NLQHBerryCurvature2015,NLHBerry1}.
The analogy of NLHE in real space can be achieved in magnetic materials
hosting topological spin textures. Recent years have witnessed the
rising of three-dimensional topological spintronics, whose central
topic is to search for the Hopfion\citep{HopifionFisher2021,Yu2023FH}, a three-dimensional
texture resembling a twisted skyrmion tube, in various of magnetic
systems, such as chiral magnets\citep{Zang2018Nanodisk,Sutcliffe2018CM,Zang2020Dynamics,Khodzhaev2022DynamicsChiral,Zheng2023Hopfion}
and frustrated magnets\citep{Sutcliffe2018Frustrated,Naya2022Frustrated}.
The previous research works\citep{NLHHopfionTatara,HopfionNLHYLiu2023}
show that $\chi_{abb}$ is proportional to the emergent toroidal moment
of the Hopfion, expressed as $\mathcal{T}_{a}^{e}=\frac{1}{2}\int d^{3}r($${\bf r}\times{\bf b})_{a}$,
where ${\bf b}$ is the emergent magnetic field originates from the
local magnetic structures through semi-classical Boltzmann equation.
It is the counterpart of Berry curvature dipole in real space. In
our work, a quantum field theoretic approach is employed to derive the NLHC
 in a three dimensional magnetic continuous model
based on Feynman diagrammatic technique. We start our calculation from
the Hamiltonian of the electrons coupling with local magnetic structure,
which is 
\begin{equation}
\mathcal{H}_{0}=-\frac{\partial_{a}^{2}}{2m}-\varepsilon_{F}+V_{imp}({\bf r})-M{\bf n}({\bf r})\cdot{\bf \sigma},
\end{equation}
where $m$ is the mass of the electrons, $\varepsilon_{F}$ is Fermi
energy and its value is determinated by the energy difference from
the bottom of an electron band with a quadratic dispersion relationship.
Here, we set $\hbar=1$. $V_{imp}({\bf r})=u_{i}\delta({\bf r}-{\bf R})$
is the impurity potential. $M=J_{sd}S$ is the rescaled strength of
$s-d$ exchange with $S$ is the length of spin field. ${\bf n}({\bf r})$
is a normalized spin vector field and ${\bf \sigma}=(\sigma^{x},\ \sigma^{y},\ \sigma^{z})$
are Pauli matrices. Einstein summation convention is employed. With a SU(2) unitary
rotation $U^{\dagger}{\bf n}\cdot{\bf \sigma}U\rightarrow\text{\ensuremath{\sigma}}^{z}$
performed, a spin gauge field $\mathcal{A}_{a}=A_{a}^{i}\frac{\sigma}{2}^{i}=-iU^{\dagger}\partial_{a}U$
emerges. Here, the upper indices for Pauli matrices are represented
by $i,\ j,\ k$ and spatial indices are always at lower position
represented by $a,b,c$ and so on. In this "rotated frame", the
spins of electrons can be up and down which are originally parallel
and antiparellel to the local magnetic orientation respectively before
rotation. The spin field can be decomposed as $\mathcal{A}={\bf A}^{\perp}\frac{\sigma}{2}^{\perp}+{\bf A}^{z}\frac{\sigma^{z}}{2}$.
The diagonal component ${\bf A}^{z}$ corresponds to the adiabatic
process where no spin flip happens and ${\bf A}^{\perp}$ describes
the non-adiabatic process since it allows spin flip. There is a
relationship between spin gauge field and magnetic structure is 
\begin{equation}
(\nabla\times{\bf A}^{z})_{a}=\varepsilon_{abc}{\bf n}\cdot(\partial_{b}{\bf n}\times\partial_{c}{\bf n})\label{eq:emer_b}
\end{equation}
and
\begin{equation}
({\bf A}_{b}^{\perp}\times{\bf A}_{c}^{\perp})^{z}=\varepsilon_{abc}{\bf n}\cdot(\partial_{b}{\bf n}\times\partial_{c}{\bf n}).
\end{equation}
The right side Eqn.(\ref{eq:emer_b}) is just the emergent magnetic
field $b_{a}$ which comes from the local magnetic structures. The
gauge field ${\bf A}^{z}$ is equivalent to the U(1) emergent gauge
field ${\bf a}$ and emergent magnetic field ${\bf b}=\nabla\times{\bf a}$
that are responsible for the topological Hall effect observed in many
skyrmionic materials\citep{Schulz2012Emergent}. With turning on the
external electromagnetic field $\partial_{a}\rightarrow\partial_{a}-ie\mathcal{A}_{Ea}$,
the new Hamiltonian is $\mathcal{H}=\mathcal{H}_{0}+\mathcal{H}_{i}=\mathcal{H}_{0}-ej_{a}\mathcal{A}_{Ea}$.
The form with more details is

\begin{equation}
\mathcal{H}=-\frac{1}{2m}(\partial_{a}-iA{}_{a}^{i}\frac{\sigma^{i}}{2}-ie\mathcal{A}_{Ea})^{2}-\varepsilon_{F}+V_{imp}-M\sigma_{z}
\end{equation}

in which $\mathcal{A}_{Ea}$ is the $U(1)$ gauge field corresponding
to the external electromagnetic fields and $e$ is the electric charge.
The coupling between electrons and gauge fields can be expressed by
the Feynman rules. The ones would be used in following calculation
are shown in Figure \ref{fig:Feynman-Rules}. The dashed tilted lines
represent the spin gauge fields. The solid tilted lines represent
the external gauge fields which correspond to external electric fields.
Different couplings are expressed by different symbols.
Through the new form of the Hamiltonian, the diagonal part is treated
as non-perturbative one which can be solved as a two-band model\citep{LHFeynmanKhono2018}.
The difference between two bands are the Fermi energies, which are
$\mu_{\sigma}=\varepsilon_{F}+\text{\ensuremath{\sigma M}}$ with
$\sigma=\pm1$ corresponding the electrons with spin \textquotedblleft up\textquotedblright{}
and ``down''. Taking consideration of impurity scattering, the retarded
(advanced) Matsubara Green's functions are expressed as 
\begin{equation}
G_{\sigma}^{R(A)}(i\omega_{n},k)=(i\omega_{n}-\varepsilon_{\mathbf{k}}+\mu+\sigma M\pm i\eta_{\sigma})^{-1}
\end{equation}
in which $\varepsilon_{k}=\frac{k^{2}}{2m}$ and $\eta_{\sigma}=\frac{1}{2\tau_{\sigma}}$.
$\tau_{\sigma}$ is the average scattering time introduced by self-energy
calculation which is $\tau_{\sigma}=\frac{1}{2\pi n_{i}u_{i}^{2}\nu_{\sigma}}$.
Moreover $n_{i}$ is the density of impurity and $\nu_{\sigma}$ is
the density-of-states of the $\sigma$ band electron at the corresponding
Fermi surface. In three dimension, the density of state is $\nu(\varepsilon)=\frac{1}{2\pi^{2}}\sqrt{\frac{m\varepsilon}{2}}$
and at Fermi surface we have $\nu_{\sigma}=\frac{1}{2\pi^{2}}\sqrt{\frac{m\mu_{\sigma}}{2}}$.
The discrete frequency $i\omega_{n}$ is replaced by a variable $z+i\delta$
for performing analytical continuation. Then the summation over all
the possible frequencies of fermions $\sum_{n}$ will be replaced
by $\frac{\beta}{2\pi i}\int dz$, in which $\beta=\frac{1}{k_{B}T}$.
Before going to the details of diagrammatic calculation, some assumptions
need to be clarified. First we assume $\tau_{\uparrow}\approx\tau_{\downarrow}=\tau$.
The relationship between $\varepsilon_{F}$ and $\tau$ is $\varepsilon_{F}\tau\gg1$
for a weak disorder situation. The contribution to the conductivities
from electrons can be devided into two parts, Fermi surface and Fermi
sea. In the weak disorder , the contribution from Fermi sea
has a factor $1/(\varepsilon_{F}\tau)$ comparing to the Fermi surface\citep{MahanBook}.
Thus, we only consider the contribution from Fermi surface here. The
Fermi momentum $k_{F}=\sqrt{2m\varepsilon_{F}}$ and Fermi velocity
is $v_{F}=\sqrt{\frac{2\varepsilon_{F}}{m}}$. The mean free path
for the electrons is $l=v_{F}\tau$.We perform the calculation at
$ql\ll1$ regime \citep{Coleman2015Book} and local effective field regime $(ql)^{2}<M\tau$
with absence of spin relexation\citep{OTN2004Hall,LHFeynmanKhono2018,WeakCouplingKhono2019},
where $q$ is the momentum of spin textures. For a
spatially smooth varying magnetic structure, the spin gauge field
is treated as pertubation. The requirement of spin gauge field is
the amplitude $|\mathcal{\mathcal{A}}(q)|\ll k_{F}$. The current
is $j_{a}=-\frac{1}{e}\frac{\delta\mathcal{H}}{\delta A_{Ea}}.$ The
second order response $\chi_{abc}$ in the response relationship has
the form
\begin{equation}
j_{a} =\sigma_{ab}E_{b}+\chi_{abc}E_{b}E_{c}+....\label{eq:response}
\end{equation}
The first term at right side of Eqn.(\ref{eq:response}) is the linear response.
The definition of second order response $\chi_{abc}$ requires that
$\chi_{abc}=\chi_{acb}$. When measuring the NHLC, the $ac$ input
voltage is employed. The external electric field has the form $E_{b}=i\omega_{b}A_{Eb}.$
The second order response shown by Eqn.(\ref{eq:response}) can be
expressed by Feynman diagrams. Inspired by the linear response theory
in magnetic systems\citep{LHBruno2004}, in which the leading order
contribution to the conductivity is linear in the density of electrons,
searching for linear $\nu_{\sigma}$ terms is the primary task in
the process of calculation. The Feynman rules are shown in Figure \ref{fig:Feynman-Rules},
where two classes of diagrams contribute to the nonlinear Hall effect,
the two-photon and triangle diagrams. Still inspired by the linear
response in magnetic systems, we make a constrain on the combination
of the the spin gauge fields will emerge in the results of NLHC. Consequentially,
diagrams may contribute to the NLHC can be further categoritzed, ones
without and with spin flip. The leading order with non-zero contribution
linear in $\nu_{\sigma}$ has been shown in Figure \ref{fig:no-switch}.
The electron flows are clockwise starting from the left vertex for
all diagrams. 
 The contribution $\mathcal{D}_{abb}(\omega,q)$ from the Figure \ref{fig:no-switch}(a)
is

\begin{align}
\mathcal{D}_{abb}^{2(a)}(\omega,q)= & \sum_{\sigma}\frac{i\sigma e^{3}}{4\pi}\int dz\int[dk]n'_{F}(z)\frac{k_{a}k_{l}A_{l}^{z}(q)}{2m^{3}}\nonumber \\
 & \times(G_{\sigma,k-\frac{q}{2}}^{R}(z+\omega)G_{\sigma,k+\frac{q}{2}}^{R}(z+\omega)G_{\sigma,k+\frac{q}{2}}^{A}(z)\nonumber \\
 & +G_{\sigma,k-\frac{q}{2}}^{R}(z)G_{\sigma,k+\frac{q}{2}}^{R}(z)G_{\sigma,k+\frac{q}{2}}^{A}(z-\omega))\label{eq:contri}
\end{align}
in which $G_{\sigma,k\pm\frac{q}{2}}^{R(A)}(z\pm\omega)=G_{\sigma}^{R(A)}(z\pm\omega,k\pm\frac{q}{2})$and
$n_{F}(z)=\frac{1}{e^{\beta(z-\mu)}+1}$ is the Fermi-Dirac distribution.
$n'_{F}(z)\approx-\delta(z)$ is employed for further simplification.
With long wave approximation, the integral can be further approximated
by making $q=0$ in the dominator.  $k_{a}k_{l}$ can be replaced
by $\frac{1}{d}\delta_{al}k^{2}$ since the electron band is isotropic
and the integral dimension is $d=3$. As a result, we have
\begin{align}
\mathcal{D}_{abb}^{2(a)}(\omega,q)\approx & \frac{e^{3}}{4\pi im^{2}}\sum_{\sigma}\int[dk]\frac{k^{2}}{3m}A_{a}^{z}(q)\nonumber \\
 & \times(G_{\sigma,k}^{R}(\omega)G_{\sigma,k}^{R}(\omega)G_{\sigma,k}^{A}(0)\nonumber \\
 & +G_{\sigma,k}^{R}(0)G_{\sigma,k}^{R}(0)G_{\sigma,k}^{A}(-\omega))
\end{align}
and similarly
\begin{align}
\mathcal{\mathcal{D}}_{abb}^{2(b)}(\omega,q)\approx & \frac{e^{3}}{4\pi im^{2}}\sum_{\sigma}\int[dk]\frac{k^{2}}{3m}A_{a}^{z}(q)\nonumber \\
 & \times(G_{\sigma,k}^{R}(\omega)G_{\sigma,k}^{A}(0)G_{\sigma,k}^{A}(0)\nonumber \\
 & +G_{\sigma,k}^{R}(0)G_{\sigma,k}^{A}(-\omega)G_{\sigma,k}^{A}(-\omega)).
\end{align}
Then the integral $\int[dk]=\int\nu(\varepsilon)d\varepsilon$ is
used for the integral and $\nu(\varepsilon)=\frac{1}{2\pi^{2}}\sqrt{\frac{m\varepsilon}{2}}$.
As a consequence, the density of states around fermi surface is $\nu_{\sigma}=\frac{1}{2\pi^{2}}\sqrt{\frac{m\mu_{\sigma}}{2}}.$
Then the contribution to the nonlinear Hall conductance under $dc$
limit is

\begin{align}
 & \Xi_{abb}^{2(a)+(b)}(q)\nonumber \\
= & \lim_{\omega\rightarrow0}\frac{\mathcal{D}_{abb}^{2(a)+(b)}(\omega,q)-\mathcal{D}_{abb}^{2(a)+(b)}(0,q)}{\omega}\nonumber \\
= & \frac{\partial\mathcal{D}_{abb}^{2(a)+(b)}(\omega,q)}{\partial\omega}|_{\omega=0}\approx\frac{e^{3}\tau^{2}A_{a}^{z}(q)}{2m^{2}}\sum_{\sigma}\sigma\nu_{\sigma}.
\end{align}
The (c) and (d) diagrams in Figure \ref{fig:no-switch} contribute
to NLHC $\chi_{aba}=\chi_{aab}$ by 
\begin{equation}
\Xi_{aba}^{2(c)+(d)}(q)=\Xi_{aab}^{2(c)+(d)}(q)\approx\frac{e^{3}\tau^{2}A_{a}^{z}(q)}{2m^{2}}\sum_{\sigma}\sigma\nu_{\sigma}.
\end{equation}
Triangle diagram contribution to $\chi_{abb}$ is shown in Figure \ref{fig:no-switch}
(e) and (f) corresponds to $\chi_{bab}$ and $\chi_{bba}$. The contribution
to the NLHC is 
\begin{align}
\Xi_{abb}^{2(e)}(q) & =\Xi_{bba}^{2(f)}(q)=\Xi_{bab}^{2(f)}(q)\approx\frac{e^{3}\tau^{2}A_{a}^{z}(q)}{2m^{2}}\sum_{\sigma}\sigma\nu_{\sigma}.
\end{align}
The diagrams with spin flip contribution to NHLC is shown in the Figure
\ref{fig:switch}.
\begin{figure}
\include{switch}

\caption{Nonzero leading order diagrams with spin flip\label{fig:switch}}
\end{figure}
 The contribution to the $dc$ nonlinear Hall conductance is

\begin{align}
 & \Xi_{abb}^{3(a)+(b)}(Q=q'-q)\nonumber \\
\approx & \frac{e^{3}\tau^{2}}{m^{2}}\frac{1}{4M^{2}\tau^{2}+1}({\bf A}_{a}^{\perp}(q')\times i\frac{\partial}{\partial q_{l}}{\bf A}_{l}^{\perp}(-q))^{z}\sum_{\sigma}\sigma\nu_{\sigma}.\label{eq:flip}
\end{align}
The details of calculation are shown in supplemental material\citep{Supp} Section II(B).
The diagrams with exchange of the left and right vertices in Figure
\ref{fig:switch} will give the contribution as
\begin{equation}
\Xi_{bab}^{3(c)+(d)}(Q)=\Xi_{bba}^{3(c)+(d)}(Q)=\Xi_{abb}^{3(a)+(b)}(Q).
\end{equation}
It is same to cases without spin flip.

\begin{figure}


\tikzset{every picture/.style={line width=0.75pt}} 

\begin{tikzpicture}[x=0.75pt,y=0.75pt,yscale=-1,xscale=1]

\draw [color={rgb, 255:red, 74; green, 144; blue, 226 }  ,draw opacity=1 ]   (64,50) -- (94,50) ;
\draw [color={rgb, 255:red, 74; green, 144; blue, 226 }  ,draw opacity=1 ] [dash pattern={on 4.5pt off 4.5pt}]  (95.5,50) -- (95.5,79.6) -- (95.5,120)(92.5,50) -- (92.5,79.6) -- (92.5,120) ;
\draw [color={rgb, 255:red, 74; green, 144; blue, 226 }  ,draw opacity=1 ]   (24,50) -- (61,50) ;
\draw [shift={(64,50)}, rotate = 180] [fill={rgb, 255:red, 74; green, 144; blue, 226 }  ,fill opacity=1 ][line width=0.08]  [draw opacity=0] (8.93,-4.29) -- (0,0) -- (8.93,4.29) -- cycle    ;
\draw [color={rgb, 255:red, 74; green, 144; blue, 226 }  ,draw opacity=1 ]   (94,50) -- (131,50) ;
\draw [shift={(134,50)}, rotate = 180] [fill={rgb, 255:red, 74; green, 144; blue, 226 }  ,fill opacity=1 ][line width=0.08]  [draw opacity=0] (8.93,-4.29) -- (0,0) -- (8.93,4.29) -- cycle    ;
\draw [color={rgb, 255:red, 74; green, 144; blue, 226 }  ,draw opacity=1 ]   (134,50) -- (164,50) ;
\draw [color={rgb, 255:red, 74; green, 144; blue, 226 }  ,draw opacity=1 ]   (127,120) -- (164,120) ;
\draw [shift={(124,120)}, rotate = 0] [fill={rgb, 255:red, 74; green, 144; blue, 226 }  ,fill opacity=1 ][line width=0.08]  [draw opacity=0] (8.93,-4.29) -- (0,0) -- (8.93,4.29) -- cycle    ;
\draw [color={rgb, 255:red, 74; green, 144; blue, 226 }  ,draw opacity=1 ]   (94,120) -- (124,120) ;
\draw [color={rgb, 255:red, 74; green, 144; blue, 226 }  ,draw opacity=1 ]   (24,120) -- (54,120) ;
\draw [color={rgb, 255:red, 74; green, 144; blue, 226 }  ,draw opacity=1 ]   (57,120) -- (94,120) ;
\draw [shift={(54,120)}, rotate = 0] [fill={rgb, 255:red, 74; green, 144; blue, 226 }  ,fill opacity=1 ][line width=0.08]  [draw opacity=0] (8.93,-4.29) -- (0,0) -- (8.93,4.29) -- cycle    ;
\draw [color={rgb, 255:red, 208; green, 2; blue, 27 }  ,draw opacity=1 ]   (244,50) -- (274,50) ;
\draw [color={rgb, 255:red, 208; green, 2; blue, 27 }  ,draw opacity=1 ] [dash pattern={on 4.5pt off 4.5pt}]  (275.5,51) -- (275.5,121)(272.5,51) -- (272.5,121) ;
\draw [color={rgb, 255:red, 208; green, 2; blue, 27 }  ,draw opacity=1 ]   (204,50) -- (241,50) ;
\draw [shift={(244,50)}, rotate = 180] [fill={rgb, 255:red, 208; green, 2; blue, 27 }  ,fill opacity=1 ][line width=0.08]  [draw opacity=0] (8.93,-4.29) -- (0,0) -- (8.93,4.29) -- cycle    ;
\draw [color={rgb, 255:red, 208; green, 2; blue, 27 }  ,draw opacity=1 ]   (274,50) -- (311,50) ;
\draw [shift={(314,50)}, rotate = 180] [fill={rgb, 255:red, 208; green, 2; blue, 27 }  ,fill opacity=1 ][line width=0.08]  [draw opacity=0] (8.93,-4.29) -- (0,0) -- (8.93,4.29) -- cycle    ;
\draw [color={rgb, 255:red, 208; green, 2; blue, 27 }  ,draw opacity=1 ]   (314,50) -- (344,50) ;
\draw [color={rgb, 255:red, 74; green, 144; blue, 226 }  ,draw opacity=1 ]   (307,120) -- (344,120) ;
\draw [shift={(304,120)}, rotate = 0] [fill={rgb, 255:red, 74; green, 144; blue, 226 }  ,fill opacity=1 ][line width=0.08]  [draw opacity=0] (8.93,-4.29) -- (0,0) -- (8.93,4.29) -- cycle    ;
\draw [color={rgb, 255:red, 74; green, 144; blue, 226 }  ,draw opacity=1 ]   (274,120) -- (304,120) ;
\draw [color={rgb, 255:red, 74; green, 144; blue, 226 }  ,draw opacity=1 ]   (204,120) -- (234,120) ;
\draw [color={rgb, 255:red, 74; green, 144; blue, 226 }  ,draw opacity=1 ]   (237,120) -- (274,120) ;
\draw [shift={(234,120)}, rotate = 0] [fill={rgb, 255:red, 74; green, 144; blue, 226 }  ,fill opacity=1 ][line width=0.08]  [draw opacity=0] (8.93,-4.29) -- (0,0) -- (8.93,4.29) -- cycle    ;

\draw (83,142) node [anchor=north west][inner sep=0.75pt]   [align=left] {(a)};
\draw (263,142) node [anchor=north west][inner sep=0.75pt]   [align=left] {(b)};
\draw (35,33.4) node [anchor=north west][inner sep=0.75pt]    {$\sigma $};
\draw (211,102.4) node [anchor=north west][inner sep=0.75pt]    {$\sigma $};
\draw (211,32.4) node [anchor=north west][inner sep=0.75pt]    {$\overline{\sigma}$};
\draw (21,71.4) node [anchor=north west][inner sep=0.75pt]    {$\Pi _{\sigma }(q,\omega)$};
\draw (195,71.4) node [anchor=north west][inner sep=0.75pt]    {$\Pi _{\overline{\sigma } \sigma }(q,\omega)$};
\draw (105,73.4) node [anchor=north west][inner sep=0.75pt]    {$q$};
\draw (291,73.4) node [anchor=north west][inner sep=0.75pt]    {$q$};

\end{tikzpicture}

\caption{Ladder corrections\label{fig:Ladder-corrections-1}}
\end{figure}
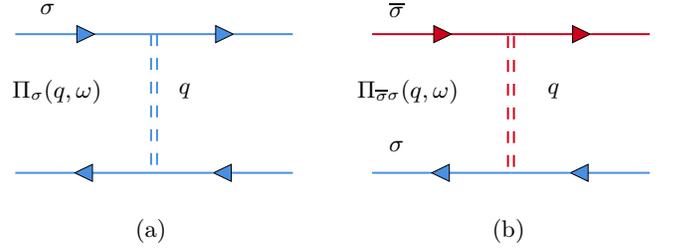

Previous research works predicted that the non-linear Hall conductance
$\chi_{abb}$ is proportional to the emergent toroidal moment which is
$\mathcal{T}_{a}^{e}=\frac{1}{2}\int[dr]({\bf r}\times{\bf B}^{z}(r))_{a}$\citep{Dubovik1990TM,Spaldin2008TM}.
Actually a general expression for the gauge field ${\bf A}^{z}(r)$
can be expressed as 
\begin{equation}
{\bf A}^{z}(r)=\frac{1}{2}{\bf B}^{z}(r)\times{\bf r}+\nabla\Lambda(r)
\end{equation}
up to a gauge transformation of the symmetric gauge. In long wave
approximation $q\rightarrow0$, the gauge field in the reciprocal
space $A_{a}^{z}(q)=\int[dr]A_{a}^{z}(r)e^{i{\bf q}\cdot{\bf r}}$
is given by 
\begin{align}
A_{a}^{z}(q=0) & =\int[dr]A_{a}^{z}(r)=-\mathcal{T}_{a}^{e}.
\end{align}
 And in real space, the density of toroidal moment can be expressed
as 
\begin{equation}
\frac{1}{2}({\bf r}\times{\bf B}^{z}(r))_{a}=r_{b}({\bf A}_{a}^{\perp}\times{\bf A}_{b}^{\perp})^{z}.
\end{equation}
In local effective regime, we have

\begin{align}
\mathcal{T}_{a}^{e} & =\int[dr]\frac{1}{2}({\bf r}\times{\bf B}^{z}(r))_{a}\nonumber \\
 & =\int[dr][dq][dq']r_{b}({\bf A}_{a}^{\perp}(q')\times{\bf A}_{b}^{\perp}(-q))^{z}e^{-i({\bf q}'-{\bf q)\cdot{\bf r}}}\nonumber \\
 & =\int[dr][dq][dq']({\bf A}_{a}^{\perp}(q')\times{\bf A}_{b}^{\perp}(-q))^{z}\frac{\partial}{i\partial q_{b}}e^{-i({\bf q'-q)\cdot{\bf r}}}\nonumber \\
 & =\int[dq]{\bf A}_{a}^{\perp}(q)\times i\frac{\partial}{\partial q_{b}}{\bf A}_{b}^{\perp}(-q).\label{eq:fourier}
\end{align}
A partial integral is performed in Eqn.(\ref{eq:fourier}). In nonlocal
effective field case , the real space expression of Eqn. (\ref{eq:flip})
would have another form\citep{LHFeynmanKhono2018,WeakCouplingKhono2019}.
Then in long wave approximation, the toroidal moment can be further
simplified as 
\begin{align}
\mathcal{T}_{a}^{e} & \approx(\int[dq])\lim_{q\rightarrow0}({\bf A}_{a}^{\perp}(q)\times i\frac{\partial}{\partial q_{b}}{\bf A}_{b}^{\perp}(-q))^{z}\nonumber \\
 & =\lim_{q\rightarrow0}({\bf A}_{a}^{\perp}(q)\times i\frac{\partial}{\partial q_{b}}{\bf A}_{b}^{\perp}(-q))^{z}.
\end{align}
To summarize, the nonlinear Hall conductivity is 
\begin{align}
\chi_{abb}^{(in)} & =\frac{1}{4}(\Xi_{abb}^{1(a)+(b)}(0)+\Xi_{abb}^{1(c)}(0)+\Xi_{abb}^{2(a)+(b)}(0))\nonumber \\
 & \approx-\frac{e^{3}\tau^{2}}{4m^{2}}(1-\frac{1}{4M^{2}\tau^{2}+1})\mathcal{T}_{a}^{e}\sum_{\sigma}\sigma\nu_{\sigma}.\label{eq:result}
\end{align}
The factor $\frac{1}{4}$ comes from the $dc$ limit approach which
is shown in supplementary material\citep{Supp} section I. And the relationships
between the nonlinear conductivities are
\begin{equation}
\frac{\chi_{aaa}^{(in)}}{2}=\chi_{abb}^{(in)}=\chi_{bab}^{(in)}=\chi_{bba}^{(in)}
\end{equation}
in which $a\neq b$. The difference between $\chi_{abb}$ and $\chi_{acc}(b\neq c)$
for an anisotropic system has not emerged in the leading order contribution.
Unlike linear response, the leading order of nonlinear Hall conductivities
at one-loop level are dependent of relaxation time $\tau$. Beyond
one-loop level, the corrections\citep{LHFeynmanKhono2018} would be
included. They can be shown by the ladder correction which are shown
in Figure \ref{fig:Ladder-corrections-1}.
 Figure \ref{fig:Ladder-corrections-1}(a) describes correction to
the spin-flip processing involved by impurtity scattering, which is
expressed as $\Pi_{\bar{\sigma}\sigma}(q)=\frac{n_{i}u_{i}^{2}(1+2i\sigma M)}{(Dq^{2}+2i\sigma M-i\omega+\tau_{s}^{-1})\tau}$.
$\tau_{s}$ is the average spin relaxation time\citep{D1971Spin}.
The contribution can be neglected under the absence of spin relaxation assumption
which is  $\tau/\tau_{s}\gg1$. That means the spin filp is too fast to encounter the impurity scattering.
Figure \ref{fig:Ladder-corrections-1}(b) contributes to vertex correction
with a factor of $\frac{Dq^{2}}{Dq^{2}-i\omega}$ where $D=\frac{1}{3}v_{F}^{2}\tau$
is a relaxation constant\citep{Coleman2015Book,LHFeynmanKhono2018}. It has no contribution
after performing $dc$ limit($\omega\rightarrow0$) and long wave
approximation($q\rightarrow0$) sequentially.

Comparing to the work using Boltzmann transport\citep{HopfionNLHYLiu2023}, 
our quantum field theoretic approach gives an additional $M\tau$ dependent result. 
Actually, strong coupling($M\tau \ll 1$) is employed in Ref.\citep{HopfionNLHYLiu2023},
 so that both results are consistent. Our result is valid for both weak coupling($M\tau<1$)
and strong coupling($M\tau>1$) regimes through long wave approximation
and local effective field assumption. On the other hand,
in weak coupling limit $M\tau\rightarrow0$ which is equivalent to the situation there is no coupling between
electrons and local magnetic textures, the electrons turn to be an isotropic
gas. $\chi_{abb}$ thus turns to be zero which is consistent with
the result of isotropic electron gas systems. 

In three dimensions, if the magnetic structure gives rise to an emergent magnetic field
as ${\bf b}=b_{0}e^{-\alpha\rho}\hat{\theta}$ along the azimuth direction in cylindrical
coordinates, where $b_{0}$ and $\alpha(\alpha>0)$ are constatns and $\rho=\sqrt{x^{2}+y^{2}}$,
the toroidal moment is given by
\begin{equation}
{\bf T}^{e}=\int[dr]\frac{1}{2}{\bf r\times{\bf b}}=\frac{\pi b_{0}}{\alpha^{2}}\hat{z}.
\end{equation}
This toroidal moment leads to a nonzero NLHC $\chi_{abb}$ in the presence of impurity scattering.
A specific example is the magnetic Hopfion. A Hopfion with Hopf number
$h=1$ can be parameterized as
\begin{equation}
{\bf n}(r)=\hat{z}+\frac{\sin2\eta(r)}{r}(x,y,0)-\frac{2\sin^{2}\eta(r)}{r^{2}}(-yz,xz,x^{2}+y^{2}),
\end{equation}
in which $r=\sqrt{x^{2}+y^{2}+z^{2}}$ and $\eta(r)$ is an arbitrary
monotonic function with constraints $\eta(0)=0$ and $\eta(\infty)=\pi$\citep{Zang2020Dynamics}. 
The emergent magnetic field can be calculated by $b_{a}=\varepsilon_{abc}{\bf n}\cdot(\partial_{b}{\bf n}\times\partial_{c}{\bf n})$.
Then the emergent toroidal moment of the Hopfion is
\begin{equation}
{\bf T}^{e}=\int[dr]\frac{1}{2}{\bf r\times{\bf b}}=\frac{2}{3}\int[dr]\sin^{2}\eta(r)\eta'(r)\hat{z}.
\end{equation}
A proper choice of $\eta(r)$ will make the toroidal moment nonzero and finite\citep{Sergey2021Scattering}. 

The origin of nonzero NLHC is diverse.  For instance, to the intrinsic NLHE, momentum space
Berry curvature dipole is the most important one.  Previous works have discussed this factor in  various of condensed matter  systems\citep{Gao2014Field,NLQHBerryCurvature2015,Liu2021Intrinsic,Wang2021Intrinsic}. Beyond intrinsic NLHE, a recent study on CuAsMn offers another mechanism for NLHE\citep{Chen2022Anti}. From a symmetry perspective,
hidden spin polarization locally breaks the inversion symmetry in real space. 
When combined with the asymmetry of electron band structures, it leads to the observable
NLHC. Our result presents another mechanism. The existence of nonzero
emergent toroidal moment breaks both the inversion and time reversal
symmetries in real space which results in a nonzero NLHC.
More generally, nonzero NHLC $\chi_{abb}$ would emerge in a variety
of multi-q magnetic states that have nonzero toroidal moment.

\

\begin{acknowledgments}
W.T.Hou was supported by the startup foundation in Tiangong University
No.63010201/52010399. J.Zang was supported by the Office of Basic
Energy Sciences, Division of Materials Sciences and Engineering, U.S.
De- partment of Energy, under Award No. DE-SC0020221. W.T.Hou would
like to thank Prof. Yi Liao and Dr. Xiaodong Ma for a beneficial discussion
about Feynam diagrammatic techniques. W.T.Hou and J.Zang would like
to thank Dr. Kazuki Nakazawa for a delighted discussion on linear
response theory in magnetic systems, and also thank Dr. Yizhou Liu
for Boltzmann transport theory. W.T.Hou would like to thank Prof. Wenhong Wang and Dr.
Jie Chen for a discussion on experiments of electronic conductivities
in magnetic systems.
\end{acknowledgments}

\bibliographystyle{unsrt}
\bibliography{nlhc}

\begin{thebibliography}{10}

\bibitem{QSHE2006}
B.A. Bernevig and S.C. Zhang.
\newblock Quantum spin hall effect.
\newblock {\em Physical Review Letters}, 96(10):106802, 2006.

\bibitem{NonlinearTRS}
Z.~Du, C.~Wang, S.~Li, H.Z. Lu, and X.~Xie.
\newblock Disorder-induced nonlinear hall effect with time-reversal symmetry.
\newblock {\em Nature Communications}, 10(1):3047, 2019.

\bibitem{NLHTRS}
C.~Ortix.
\newblock Nonlinear hall effect with time-reversal symmetry: Theory and
  material realizations.
\newblock {\em Advanced Quantum Technologies}, 4(9):2100056, 2021.

\bibitem{NLHWeylTheory2021}
R.H. Li, O.G. Heinonen, A.A. Burkov, and S.S.L. Zhang.
\newblock Nonlinear hall effect in weyl semimetals induced by chiral anomaly.
\newblock {\em Physical Review B}, 103(4):045105, 2021.

\bibitem{Machon2022Weyl}
D.G. Ovalle, A.~Pezo, and A.~Manchon.
\newblock Influence of the surface states on the nonlinear hall effect in weyl
  semimetals.
\newblock {\em Physical Review B}, 106(21):214435, 2022.

\bibitem{Yasuda2017Ti}
K.~Yasuda, A.~Tsukazaki, R.~Yoshimi, K.~Kondou, K.~S. Takahashi, Y.~Otani,
  et~al.
\newblock Current-nonlinear hall effect and spin-orbit torque magnetization
  switching in a magnetic topological insulator.
\newblock {\em Physical review letters}, 119(13):137204, 2017.

\bibitem{Rao2021Ti}
W.~Rao, Y.L. Zhou, Y.J. Wu, H.J. Duan, M.X. Deng, and R.Q. Wang.
\newblock Theory for linear and nonlinear planar hall effect in topological
  insulator thin films.
\newblock {\em Physical Review B}, 103(15):155415, 2021.

\bibitem{QiongMaWTe22019}
Q.~Ma, S.Y. Xu, H.~Shen, D.~MacNeill, V.~Fatemi, T.R. Chang, et~al.
\newblock Observation of the nonlinear hall effect under
  time-reversal-symmetric conditions.
\newblock {\em Nature}, 565(7739):337--342, 2019.

\bibitem{NLHGraphene2}
C.P. Zhang, J.~Xiao, B.T. Zhou, J.X. Hu, Y.M. Xie, B.~Yan, et~al.
\newblock Giant nonlinear hall effect in strained twisted bilayer graphene.
\newblock {\em Physical Review B}, 106(4):L041111, 2022.

\bibitem{NLHGraphene2022}
J.~Duan, Y.~Jian, Y.~Gao, H.~Peng, J.~Zhong, Q.~Feng, et~al.
\newblock Giant second-order nonlinear hall effect in twisted bilayer graphene.
\newblock {\em Physical review letters}, 129(18):186801, 2022.

\bibitem{WeylTaIrTe42021}
D.~Kumar, C.H. Hsu, R.~Sharma, T.R. Chang, P.~Yu, J.~Wang, et~al.
\newblock Room-temperature nonlinear hall effect and wireless radiofrequency
  rectification in weyl semimetal tairte4.
\newblock {\em Nature Nanotechnology}, 16(4):421--425, 2021.

\bibitem{NHAnti2023}
A.~Gao, Y.F. Liu, J.X. Qiu, B.~Ghosh, T.~V.~Trevisan, Y.~Onishi, et~al.
\newblock Quantum metric nonlinear hall effect in a topological
  antiferromagnetic heterostructure.
\newblock {\em Science}, 381(6654):181--186, 2023.

\bibitem{NLHBolz2019}
S.~Nandy and I.~Sodemann.
\newblock Symmetry and quantum kinetics of the nonlinear hall effect.
\newblock {\em Physical Review B}, 100(19):195117, 2019.

\bibitem{NLHBoltz2019}
C.~Xiao, Z.~Du, and Q.~Niu.
\newblock Theory of nonlinear hall effects: Modified semiclassics from quantum
  kinetics.
\newblock {\em Physical Review B}, 100(16):165422, 2019.

\bibitem{NLHDiagramMoore2019}
D.E. Parker, T.~Morimoto, J.~Orenstein, and J.E. Moore.
\newblock Diagrammatic approach to nonlinear optical response with application
  to weyl semimetals.
\newblock {\em Physical Review B}, 99(4):045121, 2019.

\bibitem{NLHGeneralHaizhou2021}
Z.~Du, C.~Wang, H.P. Sun, H.Z. Lu, and X.~Xie.
\newblock Quantum theory of the nonlinear hall effect.
\newblock {\em Nature Communications}, 12(1):5038, 2021.

\bibitem{NLHGeneral2}
Y.~Michishita and R.~Peters.
\newblock Effects of renormalization and non-hermiticity on nonlinear responses
  in strongly correlated electron systems.
\newblock {\em Physical Review B}, 103(19):195133, 2021.

\bibitem{Chen2024Lattice}
R.~Chen, Z.~Z. Du, H.~P Sun, H.~Z. Lu, and X.C. Xie.
\newblock Nonlinear hall effect on a disordered lattice.
\newblock {\em Physical Review B}, 110(8):L081301, 2024.

\bibitem{Gao2014Field}
Y.~Gao, S.A. Yang, and Q.~Niu.
\newblock Field induced positional shift of bloch electrons and its dynamical
  implications.
\newblock {\em Physical review letters}, 112(16):166601, 2014.

\bibitem{NLQHBerryCurvature2015}
I.~Sodemann and L.~Fu.
\newblock Quantum nonlinear hall effect induced by berry curvature dipole in
  time-reversal invariant materials.
\newblock {\em Physical Review Letters}, 115(21):216806, 2015.

\bibitem{NLHBerry1}
S.~Lai, H.~Liu, Z.~Zhang, J.~Zhao, X.~Feng, N.~Wang, et~al.
\newblock Third-order nonlinear hall effect induced by the berry-connection
  polarizability tensor.
\newblock {\em Nature Nanotechnology}, 16(8):869--873, 2021.

\bibitem{HopifionFisher2021}
N.~Kent, N.~Reynolds, D.~Raftrey, I.T. Campbell, S.~Virasawmy, S.~Dhuey, et~al.
\newblock Creation and observation of hopfions in magnetic multilayer systems.
\newblock {\em Nature Communications}, 12(1):1562, 2021.

\bibitem{Yu2023FH}
X.~Yu, Y.~Liu, K.V. Iakoubovskii, K.~Nakajima, N.~Kanazawa, N.~Nagaosa, and
  Y.~Tokura.
\newblock Realization and current-driven dynamics of fractional hopfions and
  their ensembles in a helimagnet fege.
\newblock {\em Advanced Materials}, 35(20):2210646, 2023.

\bibitem{Zang2018Nanodisk}
Y.~Liu, R.K. Lake, and J.~Zang.
\newblock Binding a hopfion in a chiral magnet nanodisk.
\newblock {\em Physical Review B}, 98(17):174437, 2018.

\bibitem{Sutcliffe2018CM}
P.~Sutcliffe.
\newblock Hopfions in chiral magnets.
\newblock {\em Journal of Physics A: Mathematical and Theoretical},
  51(37):375401, 2018.

\bibitem{Zang2020Dynamics}
Y.~Liu, W.T. Hou, X.~Han, and J.~Zang.
\newblock Three-dimensional dynamics of a magnetic hopfion driven by spin
  transfer torque.
\newblock {\em Physical Review Letters}, 124(12):127204, 2020.

\bibitem{Khodzhaev2022DynamicsChiral}
Z.~Khodzhaev and E.~Turgut.
\newblock Hopfion dynamics in chiral magnets.
\newblock {\em Journal of Physics: Condensed Matter}, 34(22):225805, 2022.

\bibitem{Zheng2023Hopfion}
F.~Zheng, N.S. Kiselev, F.N. Rybakov, L.~Yang, W.~Shi, S.~Bl{\"u}gel, et~al.
\newblock Hopfion rings in a cubic chiral magnet.
\newblock {\em Nature}, 623(7988):718--723, 2023.

\bibitem{Sutcliffe2018Frustrated}
P.~Sutcliffe.
\newblock Skyrmion knots in frustrated magnets.
\newblock {\em Physical review letters}, 118(24):247203, 2017.

\bibitem{Naya2022Frustrated}
C.~Naya, D.~Schubring, M.~Shifman, and Z.~Wang.
\newblock Skyrmions and hopfions in three-dimensional frustrated magnets.
\newblock {\em Physical Review B}, 106(9):094434, 2022.

\bibitem{NLHHopfionTatara}
B.~G{\"0}bel, C.A. Akosa, G.~Tatara, and I.~Mertig.
\newblock Topological hall signatures of magnetic hopfions.
\newblock {\em Physical Review Research}, 2(1):013315, 2020.

\bibitem{HopfionNLHYLiu2023}
Y.~Liu, H.~Watanabe, and N.~Nagaosa.
\newblock Emergent magnetomultipoles and nonlinear responses of a magnetic
  hopfion.
\newblock {\em Physical Review Letters}, 129(26):267201, 2022.

\bibitem{Schulz2012Emergent}
T.~Schulz, R.~Ritz, A.~Bauer, M.~Halder, M.~Wagner, C.~Franz, et~al.
\newblock Emergent electrodynamics of skyrmions in a chiral magnet.
\newblock {\em Nature Physics}, 8(4):301--304, 2012.

\bibitem{LHFeynmanKhono2018}
K.~Nakazawa, M.~Bibes, and H.~Kohno.
\newblock Topological hall effect from strong to weak coupling.
\newblock {\em Journal of the Physical Society of Japan}, 87(3):033705, 2018.

\bibitem{MahanBook}
G.D. Mahan.
\newblock {\em Many-particle physics}.
\newblock Springer Science {\&} Business Media, 2013.

\bibitem{Coleman2015Book}
P.~Coleman.
\newblock {\em Introduction to many-body physics}.
\newblock Cambridge University Press, 2015.

\bibitem{OTN2004Hall}
M.~Onoda, G.~Tatara, and N.~Nagaosa.
\newblock Anomalous hall effect and skyrmion number in real and momentum
  spaces.
\newblock {\em Journal of the Physical Society of Japan}, 73(10):2624--2627,
  2004.

\bibitem{WeakCouplingKhono2019}
K.~Nakazawa and H.~Kohno.
\newblock Weak coupling theory of topological hall effect.
\newblock {\em Physical Review B}, 99(17):174425, 2019.

\bibitem{LHBruno2004}
P.~Bruno, V.~Dugaev, and M.~Taillefumier.
\newblock Topological hall effect and berry phase in magnetic nanostructures.
\newblock {\em Physical Review Letters}, 93(9):096806, 2004.

\bibitem{Supp}
\url{URL_will_be_inserted_by_publisher}.

\bibitem{Dubovik1990TM}
V.M. Dubovik and V.V. Tugushev.
\newblock Toroid moments in electrodynamics and solid-state physics.
\newblock {\em Physics reports}, 187(4):145--202, 1990.

\bibitem{Spaldin2008TM}
N.A Spaldin, M.~Fiebig, and M.~Mostovoy.
\newblock The toroidal moment in condensed-matter physics and its relation to
  the magnetoelectric effect.
\newblock {\em Journal of Physics: Condensed Matter}, 20(43):434203, 2008.

\bibitem{D1971Spin}
M.I. D'yakonov and V.I. Perel.
\newblock Spin orientation of electrons associated with the interband
  absorption of light in semiconductors.
\newblock {\em Soviet Journal of Experimental and Theoretical Physics},
  33:1053, 1971.

\bibitem{Sergey2021Scattering}
S.~S. Pershoguba, D.~Andreoli, and J.~Zang.
\newblock Electronic scattering off a magnetic hopfion.
\newblock {\em Physical Review B}, 104(7):075102, 2021.

\bibitem{Liu2021Intrinsic}
H.~Liu, J.~Zhao, Y.~X. Huang, W.~Wu, X.~L. Sheng, C.~Xiao, and S.~A. Yang.
\newblock Intrinsic second-order anomalous hall effect and its application in
  compensated antiferromagnets.
\newblock {\em Physical Review Letters}, 127(27):277202, 2021.

\bibitem{Wang2021Intrinsic}
C.~Wang, Y.~Gao, and D.~Xiao.
\newblock Intrinsic nonlinear hall effect in antiferromagnetic tetragonal
  cumnas.
\newblock {\em Physical Review Letters}, 127(27):277201, 2021.

\bibitem{Chen2022Anti}
W.~Chen, M.~Gu, J.~Li, P.~Wang, and Q.~Liu.
\newblock Role of hidden spin polarization in nonreciprocal transport of
  antiferromagnets.
\newblock {\em Physical Review Letters}, 129(27):276601, 2022.

\end{thebibliography}


\begin{thebibliography}{1}

\bibitem{NLHDiagramMoore2019}
D.E. Parker, T.~Morimoto, J.~Orenstein, and J.E. Moore.
\newblock Diagrammatic approach to nonlinear optical response with application
  to weyl semimetals.
\newblock {\em Physical Review B}, 99(4):045121, 2019.

\bibitem{NLHGeneralHaizhou2021}
Z.~Du, C.~Wang, H.P. Sun, H.Z. Lu, and X.~Xie.
\newblock Quantum theory of the nonlinear hall effect.
\newblock {\em Nature Communications}, 12(1):5038, 2021.

\bibitem{NLHGeneral2}
Y.~Michishita and R.~Peters.
\newblock Effects of renormalization and non-hermiticity on nonlinear responses
  in strongly correlated electron systems.
\newblock {\em Physical Review B}, 103(19):195133, 2021.

\bibitem{NLHGeneral3}
H.~Rostami, M.I. Katsnelson, G.~Vignale, and M.~Polini.
\newblock Gauge invariance and ward identities in nonlinear response theory.
\newblock {\em Annals of Physics}, 431, 2021.

\end{thebibliography}

\end{document}


\title{Supplemental materials for “Microscopic Theory of Nonlinear Hall Effect
in Three-dimensional Magnetic Systems”}
\author{Wen-Tao Hou}
\affiliation{School of Physical Science and Technology, Tiangong University, Tianjin,300387,
China}
\author{Jiadong Zang}
\affiliation{Department of Physics and Astronomy, University of New Hampshire,
Durham, New Hampshire 03824, USA}
\maketitle

\section{General Theory For Second Order Response }

The generic formula of quantum nonlinear Hall conductivity have been
deduced in several references\citep{NLHDiagramMoore2019,NLHGeneralHaizhou2021,NLHGeneral2,NLHGeneral3}.
The general formula of response in frequency space is 
\begin{equation}
J_{a}(t)=\Pi_{ab}(\omega_{b})\mathcal{E}_{b}e^{-i\omega_{b}t}+\frac{1}{2}\Xi_{abc}(\omega_{b},\omega_{c})\mathcal{E}_{b}\mathcal{E}_{c}e^{-i(\omega_{b}+\omega_{c})t}+....
\end{equation}
The $ac$ electric field is $E_{b}(t)=\Re[\mathcal{E}_{b}e^{-i\omega_{b}t}]=\mathcal{E}_{b}\cos(\omega_{b}t)$.
So the current has the form 

\begin{align}
J_{a}(t)= & \sigma_{ab}\mathcal{E}_{b}\cos(\omega_{b}t)+\tilde{\sigma}_{ab}\mathcal{E}_{b}\sin(\omega_{b}t)+\xi_{abc}\mathcal{E}_{b}\mathcal{E}_{c}\cos[(\omega_{b}-\omega_{c})t]+\tilde{\xi}_{abc}\mathcal{E}_{b}\mathcal{E}_{c}\sin[(\omega_{b}-\omega_{c})t]\nonumber \\
 & +\chi_{abc}\mathcal{E}_{b}\mathcal{E}_{c}\cos[(\omega_{b}+\omega_{c})t]+\tilde{\chi}_{abc}\mathcal{E}_{b}\mathcal{E}_{c}\sin[(\omega_{b}+\omega_{c})t]+...
\end{align}
in which $\Re$ means real part. The ones without and with tilde represent
the dissipative and reactive responses to the input $ac$ electric
field. After performing $dc$ limit, dissipative responses survice.
So there is a relationship as 
\begin{equation}
\chi_{abc}=\frac{1}{4}\Xi_{abc}(0,\ 0).
\end{equation}
The contributions from Fermi surface are 
\begin{align}
\Xi_{abc}^{I} & =-\frac{e^{3}}{2\pi}\Im\{Tr\int[dk]\int dzn'(z)[v_{a}\frac{\partial G^{R}(z)}{\partial z}v_{b}G^{R}(z)v_{c}G^{A}(z)]+(b\leftrightarrow c)\}
\end{align}
in which $\Im$ means imaginary part and it corresponds to the triangle
diagrams. For the two-phonon diagrams, the general formula is 
\begin{align}
\Xi_{abc}^{II} & =-\frac{e^{3}}{\pi}\Im\{Tr\int[dk]\int dzn'(z)[v_{a}\frac{\partial G^{R}(z)}{\partial z}v_{bc}G^{A}(z)]+(b\leftrightarrow c)\}
\end{align}
The two derivations on $z$ in the formula above which are from $dc$
limit. Each derivation is responsible for one $\omega$ in the general formula
$j_{a}=\chi_{abc}(\omega_{b}+\omega_{c})E_{b}(\omega_{b})E_{c}(\omega_{c})$.
Then we can recover one $\omega$. We can rewrite as 
\begin{equation}
\Xi_{abc}^{I(II)}(0)=\frac{\partial\mathcal{D}_{abc}^{I(II)}(\omega)}{\partial\omega}|_{\omega=0}
\end{equation}
in which 
\begin{align}
\mathcal{D}_{abc}^{I}(\omega) & =\frac{ie^{3}}{4\pi}Tr\int dz\int[dk]n'(z)v_{a}[G^{R}(z+\omega)v_{b}G^{R}(z)v_{c}G^{A}(z)+G^{R}(z)v_{b}G^{A}(z)v_{c}G^{A}(z-\omega)]+(b\leftrightarrow c)
\end{align}
and 
\begin{align}
\mathcal{D}_{abc}^{II}(\omega) & =\frac{ie^{3}}{2\pi}Tr\int dz\int[dk]n'(z)v_{a}[G^{R}(z+\omega)v_{bc}G^{A}(z)+G^{R}(z)v_{bc}G^{A}(z-\omega)]+(b\leftrightarrow c).
\end{align}
Based on form of $\mathcal{D}$ function and Feynman rules, we can
calculate the NLHC $\chi_{abc}$. We perform the calculation in three
dimension. Previous works\citep{NLHGeneralHaizhou2021,NLHGeneral2}
have proved that terms contain $Tr(v_{abc}G)$ and $Tr(v_{ab}Gv_{c}G)$
will be canceled. In our calculation, when there is a vertex on the
edge of two-phonon diagrams, $Tr(v_{ab}G(v_{l}A_{l}^{z})v_{c}G)$
cannot be canceled. These diagrams may have nonzero contribution at
linear order of $\nu_{\sigma}$.

\section{Two-Phonon Diagrams}

\subsection{Diagrams without spin flip}

\begin{figure}


\tikzset{every picture/.style={line width=0.75pt}} 

\begin{tikzpicture}[x=0.75pt,y=0.75pt,yscale=-1,xscale=1]

\draw  [color={rgb, 255:red, 74; green, 144; blue, 226 }  ,draw opacity=1 ] (30.17,84.8) .. controls (30.17,65.91) and (50.17,50.6) .. (74.84,50.6) .. controls (99.5,50.6) and (119.5,65.91) .. (119.5,84.8) .. controls (119.5,103.69) and (99.5,119) .. (74.84,119) .. controls (50.17,119) and (30.17,103.69) .. (30.17,84.8) -- cycle ;
\draw  [color={rgb, 255:red, 74; green, 144; blue, 226 }  ,draw opacity=1 ] (235.67,84.2) .. controls (235.67,65.31) and (255.67,50) .. (280.34,50) .. controls (305,50) and (325,65.31) .. (325,84.2) .. controls (325,103.09) and (305,118.4) .. (280.34,118.4) .. controls (255.67,118.4) and (235.67,103.09) .. (235.67,84.2) -- cycle ;
\draw  [fill={rgb, 255:red, 0; green, 0; blue, 0 }  ,fill opacity=1 ] (231.17,84.11) .. controls (231.17,80.63) and (233.04,77.85) .. (235.34,77.9) .. controls (237.65,77.95) and (239.52,80.81) .. (239.52,84.29) .. controls (239.52,87.77) and (237.65,90.55) .. (235.34,90.5) .. controls (233.04,90.45) and (231.17,87.59) .. (231.17,84.11) -- cycle ;
\draw  [fill={rgb, 255:red, 0; green, 0; blue, 0 }  ,fill opacity=1 ] (25.87,84.8) .. controls (25.87,82.43) and (27.8,80.5) .. (30.17,80.5) .. controls (32.55,80.5) and (34.47,82.43) .. (34.47,84.8) .. controls (34.47,87.17) and (32.55,89.1) .. (30.17,89.1) .. controls (27.8,89.1) and (25.87,87.17) .. (25.87,84.8) -- cycle ;
\draw  [fill={rgb, 255:red, 0; green, 0; blue, 0 }  ,fill opacity=1 ] (320,79.2) -- (330,79.2) -- (330,89.2) -- (320,89.2) -- cycle ;
\draw  [color={rgb, 255:red, 74; green, 144; blue, 226 }  ,draw opacity=1 ] (34.17,235.8) .. controls (34.17,216.91) and (54.17,201.6) .. (78.84,201.6) .. controls (103.5,201.6) and (123.5,216.91) .. (123.5,235.8) .. controls (123.5,254.69) and (103.5,270) .. (78.84,270) .. controls (54.17,270) and (34.17,254.69) .. (34.17,235.8) -- cycle ;
\draw  [fill={rgb, 255:red, 0; green, 0; blue, 0 }  ,fill opacity=1 ] (30,235.71) .. controls (30,232.23) and (31.87,229.45) .. (34.17,229.5) .. controls (36.48,229.55) and (38.35,232.41) .. (38.35,235.89) .. controls (38.35,239.37) and (36.48,242.15) .. (34.17,242.1) .. controls (31.87,242.05) and (30,239.19) .. (30,235.71) -- cycle ;
\draw   (73.56,201.6) .. controls (73.56,198.29) and (75.92,195.6) .. (78.84,195.6) .. controls (81.75,195.6) and (84.12,198.29) .. (84.12,201.6) .. controls (84.12,204.91) and (81.75,207.6) .. (78.84,207.6) .. controls (75.92,207.6) and (73.56,204.91) .. (73.56,201.6) -- cycle ; \draw   (73.56,201.6) -- (84.12,201.6) ; \draw   (78.84,195.6) -- (78.84,207.6) ;
\draw  [color={rgb, 255:red, 74; green, 144; blue, 226 }  ,draw opacity=1 ] (235.67,234.2) .. controls (235.67,215.31) and (255.67,200) .. (280.34,200) .. controls (305,200) and (325,215.31) .. (325,234.2) .. controls (325,253.09) and (305,268.4) .. (280.34,268.4) .. controls (255.67,268.4) and (235.67,253.09) .. (235.67,234.2) -- cycle ;
\draw  [fill={rgb, 255:red, 0; green, 0; blue, 0 }  ,fill opacity=1 ] (231.5,234.11) .. controls (231.5,230.63) and (233.37,227.85) .. (235.67,227.9) .. controls (237.98,227.95) and (239.85,230.81) .. (239.85,234.29) .. controls (239.85,237.77) and (237.98,240.55) .. (235.67,240.5) .. controls (233.37,240.45) and (231.5,237.59) .. (231.5,234.11) -- cycle ;
\draw   (275.06,268.4) .. controls (275.06,265.09) and (277.42,262.4) .. (280.34,262.4) .. controls (283.25,262.4) and (285.62,265.09) .. (285.62,268.4) .. controls (285.62,271.71) and (283.25,274.4) .. (280.34,274.4) .. controls (277.42,274.4) and (275.06,271.71) .. (275.06,268.4) -- cycle ; \draw   (275.06,268.4) -- (285.62,268.4) ; \draw   (280.34,262.4) -- (280.34,274.4) ;
\draw  [color={rgb, 255:red, 74; green, 144; blue, 226 }  ,draw opacity=1 ] (36.5,384.8) .. controls (36.5,365.91) and (56.5,350.6) .. (81.16,350.6) .. controls (105.83,350.6) and (125.83,365.91) .. (125.83,384.8) .. controls (125.83,403.69) and (105.83,419) .. (81.16,419) .. controls (56.5,419) and (36.5,403.69) .. (36.5,384.8) -- cycle ;
\draw   (75.88,351.6) .. controls (75.88,348.29) and (78.25,345.6) .. (81.16,345.6) .. controls (84.08,345.6) and (86.44,348.29) .. (86.44,351.6) .. controls (86.44,354.91) and (84.08,357.6) .. (81.16,357.6) .. controls (78.25,357.6) and (75.88,354.91) .. (75.88,351.6) -- cycle ; \draw   (75.88,351.6) -- (86.44,351.6) ; \draw   (81.16,345.6) -- (81.16,357.6) ;
\draw  [color={rgb, 255:red, 74; green, 144; blue, 226 }  ,draw opacity=1 ] (236.5,384.2) .. controls (236.5,365.31) and (256.5,350) .. (281.16,350) .. controls (305.83,350) and (325.83,365.31) .. (325.83,384.2) .. controls (325.83,403.09) and (305.83,418.4) .. (281.16,418.4) .. controls (256.5,418.4) and (236.5,403.09) .. (236.5,384.2) -- cycle ;
\draw   (275.88,418.4) .. controls (275.88,415.09) and (278.25,412.4) .. (281.16,412.4) .. controls (284.08,412.4) and (286.44,415.09) .. (286.44,418.4) .. controls (286.44,421.71) and (284.08,424.4) .. (281.16,424.4) .. controls (278.25,424.4) and (275.88,421.71) .. (275.88,418.4) -- cycle ; \draw   (275.88,418.4) -- (286.44,418.4) ; \draw   (281.16,412.4) -- (281.16,424.4) ;
\draw  [fill={rgb, 255:red, 0; green, 0; blue, 0 }  ,fill opacity=1 ] (121.65,384.71) .. controls (121.65,381.23) and (123.52,378.45) .. (125.83,378.5) .. controls (128.13,378.55) and (130,381.41) .. (130,384.89) .. controls (130,388.37) and (128.13,391.15) .. (125.83,391.1) .. controls (123.52,391.05) and (121.65,388.19) .. (121.65,384.71) -- cycle ;
\draw  [fill={rgb, 255:red, 0; green, 0; blue, 0 }  ,fill opacity=1 ] (321.65,384.11) .. controls (321.65,380.63) and (323.52,377.85) .. (325.83,377.9) .. controls (328.13,377.95) and (330,380.81) .. (330,384.29) .. controls (330,387.77) and (328.13,390.55) .. (325.83,390.5) .. controls (323.52,390.45) and (321.65,387.59) .. (321.65,384.11) -- cycle ;
\draw  [fill={rgb, 255:red, 0; green, 0; blue, 0 }  ,fill opacity=1 ] (114.5,79.8) -- (124.5,79.8) -- (124.5,89.8) -- (114.5,89.8) -- cycle ;
\draw  [fill={rgb, 255:red, 0; green, 0; blue, 0 }  ,fill opacity=1 ] (118.5,230.8) -- (128.5,230.8) -- (128.5,240.8) -- (118.5,240.8) -- cycle ;
\draw  [fill={rgb, 255:red, 0; green, 0; blue, 0 }  ,fill opacity=1 ] (320,229.2) -- (330,229.2) -- (330,239.2) -- (320,239.2) -- cycle ;
\draw  [fill={rgb, 255:red, 0; green, 0; blue, 0 }  ,fill opacity=1 ] (31.5,379.8) -- (41.5,379.8) -- (41.5,389.8) -- (31.5,389.8) -- cycle ;
\draw  [fill={rgb, 255:red, 0; green, 0; blue, 0 }  ,fill opacity=1 ] (231.5,379.2) -- (241.5,379.2) -- (241.5,389.2) -- (231.5,389.2) -- cycle ;

\draw (69,132) node [anchor=north west][inner sep=0.75pt]   [align=left] {(a)};
\draw (-2,76) node [anchor=north west][inner sep=0.75pt]    {$ \begin{array}{l}
ev_{a} \ 
\end{array}$};
\draw (269,132) node [anchor=north west][inner sep=0.75pt]   [align=left] {(b)};
\draw (190,72) node [anchor=north west][inner sep=0.75pt]    {$e\frac{A_{a}^{z}}{m}\frac{\sigma }{2}$};
\draw (127,72) node [anchor=north west][inner sep=0.75pt]    {$\frac{e^{2}}{2m} \delta _{bc}$};
\draw (71,291) node [anchor=north west][inner sep=0.75pt]   [align=left] {(c)};
\draw (271,291) node [anchor=north west][inner sep=0.75pt]   [align=left] {(d)};
\draw (69.33,441) node [anchor=north west][inner sep=0.75pt]   [align=left] {(e)};
\draw (271,441) node [anchor=north west][inner sep=0.75pt]   [align=left] {(f)};
\draw (56,178) node [anchor=north west][inner sep=0.75pt]    {$v_{l} A_{l}^{z}\frac{\ \sigma }{2}$};

\end{tikzpicture}

\caption{Two-phonon diagrams without spin flip\label{fig:two-phonon1}}
\end{figure}
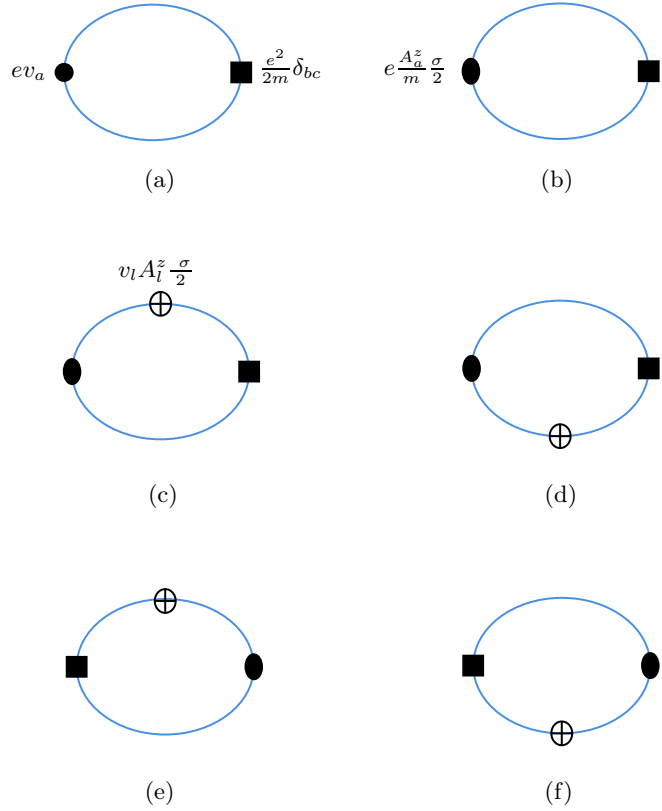

Besides the diagrams in the main text, the two-phonon diagrams may
contribute to $\chi_{abb}$ at the order $\frac{e^{3}}{m^{2}}$ are
shown in Figure \ref{fig:two-phonon1}. It is easily to prove that
the distribution of diagram (a) is zero because $\mathcal{D}_{abc}^{(a)}(q=0)\propto\int[dk]\frac{k_{a}}{N(k)}$
with $N(k)=N(-k).$ Then we have
 
\begin{align}
\Xi_{abb}^{0(b)}(q)\approx & -\frac{e^{3}}{2\pi}\sum_{\sigma}\frac{\sigma A_{a}^{z}(q)}{4m^{2}}\Im\int dzn'_{F}(z)\int[dk]\frac{\partial G_{\sigma}^{R}(\varepsilon_{k})}{\partial\varepsilon_{k}}G^{A}(\varepsilon)\nonumber \\
= & \sum_{\sigma}\frac{e^{3}\sigma A_{a}^{z}(q)}{8\pi m^{2}}\Im\int_{0}^{\infty}d\varepsilon\frac{C\sqrt{\varepsilon_{k}}}{(\varepsilon_{k}-\mu_{\sigma}-i\eta)^{2}(\varepsilon_{k}-\mu_{\sigma}+i\eta)}.\label{eq:no-insert}
\end{align}
Then we find leading order of Eqn.(\ref{eq:no-insert}) is at $\frac{e^{3}}{m^{2}}\nu_{\sigma}^{-\frac{3}{2}}$
under the condition $\varepsilon_{F}\tau\gg1$. Diagrams(e)\textasciitilde (f)
are other diagrams with one more vertex on the edge besides the ones
in the main text and they have the order $(A^{z})^{2}$. With the
assumption of smooth varying magnetic structures, it is a small quantity
comparing to the $A_{a}^{z}$ and ${\bf A}_{a}^{\perp}\times{\bf A}_{b}^{\perp}$,
we won't consider this contribution here.

\subsection{Diagrams with spin flip}

The process is shown in Figure 3 in the main text. And the contribution can be written as 
\begin{align}
 & \mathcal{D}_{abb}^{1'(a)+(b)}(\omega,q'-q)\nonumber \\
= & \frac{ie^{3}}{4\pi}\frac{1}{8m^{3}}\int dzn'_{F}(z)\int[dk]\sum_{\sigma,l\neq a}i\sigma({\bf A}_{a}^{\perp}(q')\times{\bf A}_{l}^{\perp}(-q))^{z}\nonumber \\
 & \times[(2k_{l}+q_{l})(G_{\bar{\sigma},k+q}^{R}(z+\omega)G_{\sigma,k}^{R}(z+\omega)G_{\sigma,k}^{A}(z)+G_{\bar{\sigma},k+q}^{R}(z)G_{\sigma,k}^{R}(z)G_{\sigma,k}^{A}(z-\omega))\nonumber \\
 & +(2k_{l}-q_{l})(G_{\bar{\sigma},k}^{R}(z+\omega)G_{\bar{\sigma},k}^{A}(z)G_{\sigma,k-q}^{A}(z)+G_{\bar{\sigma},k}^{R}(z)G_{\bar{\sigma},k}^{A}(z-\omega)G_{\sigma.k-q}^{A}(z-\omega))]\nonumber \\
= & \frac{e^{3}}{4\pi}\sum_{\sigma}\frac{\sigma}{4m^{3}}\int[dk]({\bf A}_{a}^{\perp}(q')\times{\bf A}_{l}^{\perp}(-q))^{z}\nonumber \\
 & \times[(2k+q)_{l}(G_{\bar{\sigma},k+q}^{R}(\omega)G_{\sigma,k}^{R}(\omega)G_{\sigma}^{A}+G_{\bar{\sigma}+}^{R}G_{\sigma}^{R}G_{\sigma,k}^{A}(-\omega))\nonumber \\
 & +(2k-q)_{l}\times(G_{\bar{\sigma},k}^{R}(\omega)G_{\bar{\sigma}}^{A}G_{\sigma-}^{A}+G_{\bar{\sigma}}^{R}G_{\bar{\sigma},k}^{A}(-\omega)G_{\sigma,k-q}^{A}(-\omega))].
\end{align}
in which $G_{\sigma}^{R(A)}=G_{\sigma,k}^{R(A)}(z=0)$ and $G_{\sigma\pm}^{R(A)}=G_{\sigma,k\pm q}^{R(A)}(z=0)$.
Then we have 
\begin{align}
 & \Xi_{abb}^{1'(a)+(b)}(q'-q)\nonumber \\
= & \frac{\partial\mathcal{D}_{abb}^{1'(a)+(b)}(\omega)}{\partial\omega}|_{\omega=0}\nonumber \\
= & (\frac{e}{2m})^{3}\sum_{\sigma}\frac{\sigma}{\pi}({\bf A}_{a}^{\perp}(q')\times{\bf A}_{l}^{\perp}(-q))^{z}\nonumber \\
 & \times\frac{\partial}{\partial(i\eta)}\int[dk][(2k+q)_{l}(G_{\bar{\sigma}+}^{R}G_{\sigma}^{R}G_{\sigma}^{A})+(2k-q)_{l}(G_{\bar{\sigma}}^{R}G_{\bar{\sigma}}^{A}G_{\sigma-}^{A})]\nonumber \\
= & (\frac{e}{2m})^{3}\sum_{\sigma}\frac{\sigma}{\pi}({\bf A}_{a}^{\perp}(q')\times{\bf A}_{l}^{\perp}(-q))^{z}\frac{\partial}{\partial(i\eta)}\nonumber \\
 & \times\int[dk][(2k+q)_{l}(G_{\bar{\sigma}+}^{R}G_{\sigma}^{R}G_{\sigma}^{A})-(-2k-q)_{l}(G_{\sigma}^{R}G_{\sigma}^{A}G_{\bar{\sigma}+}^{A})]
\end{align}
Here, we use relationship $\frac{\partial G_{\sigma}^{R}(z+\omega)}{\partial\omega}|_{\omega=0}=\frac{\partial G_{\sigma}^{R}(z)}{\partial(i\eta)}$
and $\frac{\partial G^{A}(z-\omega)}{\partial\omega}|_{\omega=0}=\frac{\partial G^{A}(z)}{\partial(i\eta)}$
to simplify the calculation. We pick up

\begin{align}
A_{1}= & \frac{\partial}{\partial(i\eta)}\int[dk]2(k_{l}+q_{l})(G_{\bar{\sigma}+}^{R}+G_{\bar{\sigma}+}^{A})G_{\sigma}^{R}G_{\sigma}^{A}\nonumber \\
= & 2m\frac{\partial^{2}}{\partial(i\eta)\partial q_{l}}\int[dk]\ln[(\varepsilon_{k+q}-\mu_{\bar{\sigma}})^{2}+\eta^{2}]G_{\sigma}^{R}G_{\sigma}^{A}\nonumber \\
= & 2m\frac{\partial}{\partial q_{l}}A_{11}.
\end{align}
And the residue is 
\begin{align}
A_{2}=-\frac{\partial}{\partial(i\eta)} & \int[dk]q_{l}(G_{\bar{\sigma}+}^{R}+G_{\bar{\sigma}+}^{A})G_{\sigma}^{R}G_{\sigma}^{A}.
\end{align}
The leading order of $A_{2}$ is in $\nu_{\sigma}^{-\frac{3}{2}}$
order under the assuption $\varepsilon_{F}\tau\gg1.$ Then we have
\begin{align}
A_{11} & \approx\frac{\partial}{\partial(i\eta)}\int[dk]\ln(-\varepsilon+\mu_{\bar{\sigma}}+i\eta)G_{\sigma}^{R}G_{\sigma}^{A}\nonumber \\
 & =C\frac{\partial}{\partial(i\eta)}\int d\varepsilon\sqrt{\varepsilon}\ln(-\varepsilon+\mu_{\bar{\sigma}}+i\eta)G_{\sigma}^{R}G_{\sigma}^{A}\nonumber \\
 & \approx-\frac{\pi}{2}\frac{\nu_{\sigma}}{(\sigma M-i\eta)\eta}.
\end{align}
in which we just keep linear term of $\nu_{\sigma}.$ So the counter part of $A_{11}$ is 
\begin{align}
B_{11} & \approx\frac{\partial}{\partial(i\eta)}\int[dk]\ln(-\varepsilon+\mu_{\sigma}-i\eta)G_{\bar{\sigma}}^{R}G_{\bar{\sigma}}^{A}\nonumber \\
 & =C\frac{\partial}{\partial(i\eta)}\int d\varepsilon\sqrt{\varepsilon}\ln(-\varepsilon+\mu_{\sigma}-i\eta)G_{\bar{\sigma}}^{R}G_{\bar{\sigma}}^{A}\nonumber \\
 & \approx\frac{\pi}{2}\frac{\nu_{\bar{\sigma}}}{(\sigma M-i\eta)\eta}.
\end{align}
$A_{11}$ and $B_{11}$ should be added together, then we have 
\begin{equation}
A_{11}+B_{11}=-\frac{\pi}{2}\frac{\nu_{\sigma}-\nu_{\bar{\sigma}}}{\sigma M-i\eta}.
\end{equation}
Finally, without vertex correction, we have 
\begin{align}
 & \Xi_{abb}^{1'(a)+(b)}(q'-q)\nonumber \\
\approx & \frac{e^{3}}{8m^{2}\eta}\frac{\partial}{\partial q_{l}}F_{al}(q'-q)\sum_{\sigma}\frac{\sigma(\nu_{\sigma}-\nu_{\bar{\sigma}})}{\sigma M-i\eta}\nonumber \\
= & \frac{ie^{3}}{4m^{2}\eta}\frac{\partial}{\partial q_{l}}F_{al}(q'-q)\frac{\eta}{M^{2}+\eta^{2}}(\nu_{\uparrow}-\nu_{\downarrow})\nonumber \\
= & \frac{ie^{3}\tau^{2}}{m^{2}(4M^{2}\tau^{2}+1)}\frac{\partial}{\partial q_{l}}F_{al}(q'-q)\sum_{\sigma}\sigma\nu_{\sigma},
\end{align}
where we define $F_{al}(q'-q)=({\bf A}_{a}^{\perp}(q')\times{\bf A}_{l}^{\perp}(-q))^{z}$.
Here we perform a specific trick when calculating the processes involving the vertices
with momentum $k_{i}$. These terms can be transformed from $k_{i}(G_{\sigma(\bar{\sigma})}^{R(A)})^{n}$ into
$m\frac{\partial}{\partial k_{i}}(G_{\sigma(\bar{\sigma})}^{R(A)})^{n-1}$
by partition integral. Then we can transform $\frac{\partial}{\partial k_{i}}$
to $\frac{\partial}{\partial q_{i}}$ in which $q$ is the momentum
of the spin gauge fields in our description. For a general case, the
response function can be expressed as

\begin{align}
 & \chi_{a_{1}a_{2}...a_{n}}(\omega,q)\nonumber \\
= & \int[dk]M_{a_{1}a_{2}...a_{n-1}}(-q)N(q,k)\nonumber \\
= & \int[dk]M_{a_{1}a_{2}...a_{n-1}}(-q)[\frac{\partial}{\partial q}(G...G)(\omega,q_{n},k)]
\end{align}
and the Hamiltonian can be written as 
\begin{align}
H & =\sum_{\omega,\omega_{i}}\prod_{q_{i}}\int[dq][dq_{i}]\chi_{a_{1}a_{2}...a_{n}}(\omega,q)A_{a_{1}}(\omega_{1},q_{1})A_{a_{2}}(\omega_{2},q_{2})\nonumber \\
 & \ \ \times...A_{a_{n}}(\omega_{n},\ q_{n})\delta({\bf q}-\sum_{i}{\bf q_{i}})\delta(\omega-\sum_{i}\omega_{i}).
\end{align}
With the ${\bf A}_{i}=i\omega_{i}{\bf E}_{i}$, the Hamiltonian can
be written as

\begin{align}
H & =i^{n}\prod_{i}\omega_{i}\int[dk][dq]M_{a_{1}a_{2}...a_{n-1}}(-q)\frac{\partial}{\partial q_{n}}(G...G)(\omega,q,k)E_{a_{1}}E_{a_{2}}...E_{a_{n}}\nonumber \\
 & \approx-i^{n}(\prod_{i}\omega_{i})E_{a_{1}}E_{a_{2}}...E_{a_{n}}\int[dk][dq](G...G)(\omega,q,k)\frac{\partial}{\partial q_{n}}M_{a_{1}a_{2}...a_{n-1}}(-q)\nonumber \\
 & =i^{n}(\prod_{i}\omega_{i})E_{a_{1}}E_{a_{2}}...E_{a_{n}}\int[dq]\frac{\partial}{\partial q}(M_{a_{1}a_{2}...a_{n-1}}(-q))\int[dk](G...G)(\omega,q,k)\delta(\omega).
\end{align}
in which $G$ is the short form for the Green's function. So the corresponding function is
\begin{equation}
\chi_{a_{1}a_{2}...a_{n}}(\omega,q)=-\frac{\partial M_{a_{1}a_{2}...a_{n}}(-q)}{\partial q}\int[dk](G...G)(\omega,q,k)
\end{equation}
by partition integral based on the invariance of the Hamiltonian.

\subsection{Trigangle Diagrams}

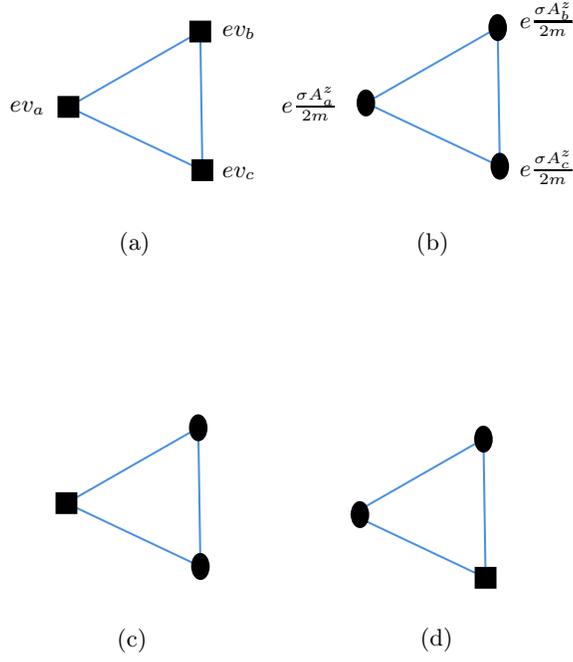
\begin{figure}


\tikzset{every picture/.style={line width=0.75pt}} 

\begin{tikzpicture}[x=0.75pt,y=0.75pt,yscale=-1,xscale=1]

\draw  [color={rgb, 255:red, 74; green, 144; blue, 226 }  ,draw opacity=1 ] (39,79.33) -- (105.47,41.3) -- (106.55,111.29) -- cycle ;
\draw  [fill={rgb, 255:red, 0; green, 0; blue, 0 }  ,fill opacity=1 ] (34,74.33) -- (44,74.33) -- (44,84.33) -- (34,84.33) -- cycle ;
\draw  [color={rgb, 255:red, 74; green, 144; blue, 226 }  ,draw opacity=1 ] (189,77.33) -- (255.47,39.3) -- (256.55,109.29) -- cycle ;
\draw  [fill={rgb, 255:red, 0; green, 0; blue, 0 }  ,fill opacity=1 ] (184.83,77.24) .. controls (184.83,73.76) and (186.7,70.98) .. (189,71.03) .. controls (191.3,71.07) and (193.17,73.93) .. (193.17,77.41) .. controls (193.17,80.89) and (191.3,83.68) .. (189,83.63) .. controls (186.7,83.58) and (184.83,80.72) .. (184.83,77.24) -- cycle ;
\draw  [color={rgb, 255:red, 74; green, 144; blue, 226 }  ,draw opacity=1 ] (38,279.33) -- (104.47,241.3) -- (105.55,311.29) -- cycle ;
\draw  [fill={rgb, 255:red, 0; green, 0; blue, 0 }  ,fill opacity=1 ] (100.3,241.21) .. controls (100.3,237.73) and (102.17,234.95) .. (104.47,235) .. controls (106.78,235.05) and (108.64,237.91) .. (108.64,241.39) .. controls (108.64,244.87) and (106.78,247.65) .. (104.47,247.6) .. controls (102.17,247.55) and (100.3,244.69) .. (100.3,241.21) -- cycle ;
\draw  [fill={rgb, 255:red, 0; green, 0; blue, 0 }  ,fill opacity=1 ] (100.47,36.3) -- (110.47,36.3) -- (110.47,46.3) -- (100.47,46.3) -- cycle ;
\draw  [fill={rgb, 255:red, 0; green, 0; blue, 0 }  ,fill opacity=1 ] (101.55,106.29) -- (111.55,106.29) -- (111.55,116.29) -- (101.55,116.29) -- cycle ;
\draw  [fill={rgb, 255:red, 0; green, 0; blue, 0 }  ,fill opacity=1 ] (251.3,39.21) .. controls (251.3,35.73) and (253.17,32.95) .. (255.47,33) .. controls (257.78,33.05) and (259.64,35.91) .. (259.64,39.39) .. controls (259.64,42.87) and (257.78,45.65) .. (255.47,45.6) .. controls (253.17,45.55) and (251.3,42.69) .. (251.3,39.21) -- cycle ;
\draw  [fill={rgb, 255:red, 0; green, 0; blue, 0 }  ,fill opacity=1 ] (252.37,109.21) .. controls (252.37,105.73) and (254.24,102.94) .. (256.55,102.99) .. controls (258.85,103.04) and (260.72,105.9) .. (260.72,109.38) .. controls (260.72,112.86) and (258.85,115.64) .. (256.55,115.59) .. controls (254.24,115.55) and (252.37,112.68) .. (252.37,109.21) -- cycle ;
\draw  [fill={rgb, 255:red, 0; green, 0; blue, 0 }  ,fill opacity=1 ] (101.37,311.21) .. controls (101.37,307.73) and (103.24,304.94) .. (105.55,304.99) .. controls (107.85,305.04) and (109.72,307.9) .. (109.72,311.38) .. controls (109.72,314.86) and (107.85,317.64) .. (105.55,317.59) .. controls (103.24,317.55) and (101.37,314.68) .. (101.37,311.21) -- cycle ;
\draw  [fill={rgb, 255:red, 0; green, 0; blue, 0 }  ,fill opacity=1 ] (33,274.33) -- (43,274.33) -- (43,284.33) -- (33,284.33) -- cycle ;
\draw  [color={rgb, 255:red, 74; green, 144; blue, 226 }  ,draw opacity=1 ] (181.83,285.03) -- (248.3,247.01) -- (249.37,317) -- cycle ;
\draw  [fill={rgb, 255:red, 0; green, 0; blue, 0 }  ,fill opacity=1 ] (244.37,312) -- (254.37,312) -- (254.37,322) -- (244.37,322) -- cycle ;
\draw  [fill={rgb, 255:red, 0; green, 0; blue, 0 }  ,fill opacity=1 ] (244.12,246.92) .. controls (244.12,243.44) and (245.99,240.66) .. (248.3,240.71) .. controls (250.6,240.76) and (252.47,243.62) .. (252.47,247.1) .. controls (252.47,250.58) and (250.6,253.36) .. (248.3,253.31) .. controls (245.99,253.26) and (244.12,250.4) .. (244.12,246.92) -- cycle ;
\draw  [fill={rgb, 255:red, 0; green, 0; blue, 0 }  ,fill opacity=1 ] (181.83,285.03) .. controls (181.83,281.55) and (183.7,278.77) .. (186,278.82) .. controls (188.3,278.87) and (190.17,281.73) .. (190.17,285.21) .. controls (190.17,288.69) and (188.3,291.47) .. (186,291.42) .. controls (183.7,291.37) and (181.83,288.51) .. (181.83,285.03) -- cycle ;

\draw (63,141) node [anchor=north west][inner sep=0.75pt]   [align=left] {(a)};
\draw (8,75) node [anchor=north west][inner sep=0.75pt]    {$ev_{a}$};
\draw (212.83,141) node [anchor=north west][inner sep=0.75pt]   [align=left] {(b)};
\draw (145,68) node [anchor=north west][inner sep=0.75pt]    {$e\frac{\sigma A_{a}^{z}}{2m}$};
\draw (62,341) node [anchor=north west][inner sep=0.75pt]   [align=left] {(c)};
\draw (214.83,340.6) node [anchor=north west][inner sep=0.75pt]   [align=left] {(d)};
\draw (115,37) node [anchor=north west][inner sep=0.75pt]    {$ev_{b}$};
\draw (115,108) node [anchor=north west][inner sep=0.75pt]    {$ev_{c}$};
\draw (265,25) node [anchor=north west][inner sep=0.75pt]    {$e\frac{\sigma A_{b}^{z}}{2m}$};
\draw (265,100) node [anchor=north west][inner sep=0.75pt]    {$e\frac{\sigma A_{c}^{z}}{2m}$};

\end{tikzpicture}


\caption{Other triangle diagrams without spin flip \label{fig:no_switch1}}
\end{figure}
\begin{figure}
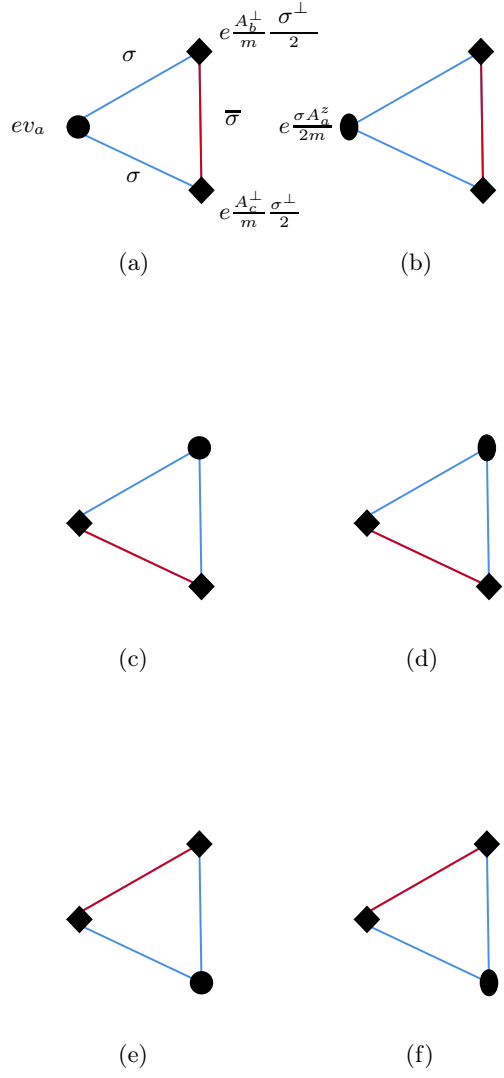

\include{triangle_with_switch}

\caption{Tiangle diagrams with spin switch flip \label{fig:Tiangle-diagrams-with}}
\end{figure}

The diagrams without spin flip have not been mentioned in the main
text are shown in Figure \ref{fig:no_switch1}. Diagam(a) will have $\mathcal{D}_{abc}\propto\int[dk]\frac{k_{a}k_{b}k_{c}}{N(k)}$
which will give a zero contribution after long wave approximation.
Others are with order of $A^{z}$ higher than one. The diagrams with
spin flip are shown in Figure \ref{fig:Tiangle-diagrams-with}.The contribution of diagrams (a) and (b) can be simply
proved to be zero owing to the exchange of indices.
The contributions from other diagrams are 

\begin{align}
 & \mathcal{D}_{abc}^{2'(c)+(e)}(\omega,Q=q'-q)\nonumber \\
= & \frac{ie^{3}}{2\pi}\frac{1}{4m^{3}}\sum_{\sigma}\int[dk]\int dzn'_{F}(z)i\sigma({\bf A}_{a}^{\perp}(q')\times{\bf A}_{b}^{\perp}(-q))^{z}k_{c}\nonumber \\
\ \  & \times(G_{\bar{\sigma},k+q}^{R}(z+\omega)G_{\sigma,k}^{R}(z)G_{\sigma,k}^{A}(z)+G_{\bar{\sigma},k+q}^{R}(z)G_{\sigma,k}^{A}(z)G_{\sigma,k}^{A}(z-\omega)\nonumber \\
 & +G_{\bar{\sigma},k}^{R}(z+\omega)G_{\bar{\sigma}}^{R}(z)G_{\sigma,k-q}^{A}(z)+G_{\bar{\sigma},k}^{R}(z)G_{\bar{\sigma},k}^{A}(z)G_{\sigma,k-q}^{A}(z-\omega))+(b\leftrightarrow c)\nonumber \\
= & \frac{e^{3}}{2\pi}\sum_{\sigma}\frac{\sigma}{4m^{3}}\int[dk]F_{al}(q',-q)k_{c}(G_{\bar{\sigma},k+q}^{R}(\omega)G_{\sigma}^{R}G_{\sigma}^{A}+G_{\bar{\sigma}+}^{R}G_{\sigma}^{A}G_{\sigma,k}^{A}(\omega)\nonumber \\
 & +G_{\bar{\sigma},k}^{R}(\omega)G_{\bar{\sigma}}^{R}G_{\sigma-}^{A}+G_{\bar{\sigma}}^{R}G_{\bar{\sigma}}^{A}G_{\sigma,k-q}^{A}(-\omega))+(b\leftrightarrow c)
\end{align}
and 
\begin{align}
 & \Xi_{abc}^{2'(c)+(e)}(Q=q'-q)\nonumber \\
= & \frac{\partial\mathcal{D}_{abc}^{2'(c)+(e)}(\omega)}{\partial\omega}|_{\omega=0}\nonumber \\
= & -\frac{e^{3}}{2\pi}\sum_{\sigma}\frac{\sigma}{4m^{3}}\int[dk]k_{c}F_{al}(q',-q)(G_{\bar{\sigma}+}^{R2}G_{\sigma}^{R}G_{\text{\ensuremath{\sigma}}}^{A}-G_{\bar{\sigma}+}^{R}G_{\sigma}^{A3}+G_{\bar{\sigma}}^{R3}G_{\sigma-}^{A}-G_{\bar{\sigma}}^{R}G_{\bar{\sigma}}^{A}G_{\sigma-}^{A2})+(b\leftrightarrow c).
\end{align}
Then we calculate the two pairs of terms from $\Xi_{abc}^{2'(c)+(e)}(Q=q'-q)$ as 
\begin{align}
 & \int[dk]F_{ab}(q,-q)k_{c}(G_{\bar{\sigma}}^{R3}G_{\sigma-}^{A}-G_{\bar{\sigma}+}^{R}G_{\sigma}^{A3})\nonumber \\
= & \frac{m}{2}\int[dk]F_{ab}(q,-q)(\frac{\partial G_{\bar{\sigma}}^{R2}}{\partial k_{c}}G_{\sigma-}^{A}-G_{\bar{\sigma}+}^{R}\frac{\partial G_{\sigma}^{A2}}{\partial k_{c}})\nonumber \\
= & \frac{m}{2}\int[dk]F_{ab}(q,-q)(G_{\bar{\sigma}}^{R2}\frac{\partial G_{\sigma-}^{A}}{\partial q_{c}}+\frac{\partial G_{\bar{\sigma}+}^{R}}{\partial q_{c}}G_{\sigma}^{A2})\nonumber \\
\rightarrow & -\frac{m}{2}\frac{\partial}{\partial q_{c}}F_{ab}(q,-q)\int[dk](G_{\bar{\sigma}}^{R2}G_{\sigma-}^{A}+G_{\bar{\sigma}+}^{R}G_{\sigma}^{A2})
\end{align}
and 
\begin{align}
 & \int[dk](G_{\bar{\sigma}}^{R2}G_{\sigma}^{A}+G_{\bar{\sigma}}^{R}G_{\sigma}^{A2})\nonumber \\
= & -\frac{C\pi i}{4}[-\frac{\sqrt{\mu_{\bar{\sigma}}+i\eta}}{(\sigma M-i\eta)^{2}}+\frac{\sqrt{\mu_{\sigma}-i\eta}}{(\sigma M-i\eta)^{2}}-\frac{1}{2\sqrt{\mu_{\bar{\sigma}}+i\eta}(\sigma M-i\eta)}-\frac{\sqrt{\mu_{\sigma}-i\eta}}{(\sigma M-i\eta)}\nonumber \\
 & +\frac{\sqrt{\mu_{\bar{\sigma}}+i\eta}}{(\sigma M-i\eta)^{2}}+\frac{1}{2\sqrt{\mu_{\sigma}-i\eta}(\sigma M-i\eta)}]
\end{align}
is zero at linear- $\nu_{\sigma}$ order. And next is $\frac{1}{\eta}\int[dk]\arctan(\frac{\varepsilon-\mu_{\sigma}}{\eta})G_{\bar{\sigma}+}^{R2}+\arctan(\frac{\varepsilon-\mu_{\bar{\sigma}}}{\eta})G_{\sigma-}^{A2}.$
Another way is to calulate 
\begin{align}
 & \int[dk]\arctan(\frac{\varepsilon-\mu_{\sigma}}{\eta})(G_{\bar{\sigma}}^{R2}-G_{\bar{\sigma}}^{A2})\nonumber \\
= & \frac{Ci}{2}\int_{0}^{\infty}d\varepsilon\sqrt{\varepsilon}\ln(\frac{1+\frac{\varepsilon-\mu_{\sigma}}{i\eta}}{1-\frac{\varepsilon-\mu_{\sigma}}{i\eta}})(G_{\bar{\sigma}}^{R2}-G_{\bar{\sigma}}^{A2})\nonumber \\
= & \frac{Ci}{2}\int d\varepsilon\sqrt{\varepsilon}\ln(\frac{\varepsilon-\mu_{\sigma}+i\eta}{-\varepsilon+\mu_{\sigma}+i\eta})(G_{\bar{\sigma}}^{R2}-G_{\bar{\sigma}}^{A2})\nonumber \\
\approx & \frac{Ci}{2}\int d\varepsilon\sqrt{\varepsilon}[\ln(-\varepsilon+\mu_{\sigma}-i\eta)-\ln(-\varepsilon+\mu_{\sigma}+i\eta_{\sigma})](G_{\bar{\sigma}}^{R2}-G_{\bar{\sigma}}^{A2})]\nonumber \\
\approx & \frac{\pi\sigma M}{2}(\nu_{\sigma}-\nu_{\bar{\sigma}})(\frac{1}{M^{2}+\eta^{2}}-\frac{1}{M^{2}}).
\end{align}
It is easy to prove the contribution is zero after summation of $\sigma$
at linear-$\nu_{\sigma}$ order, 
\begin{align}
 & \mathcal{D}_{abc}^{2(d)}(Q',\omega)\nonumber \\
\approx & \sum_{\sigma}\frac{ie^{3}\sigma}{16\pi m^{3}}\int dzn'_{F}(z)i\sigma\int[dk]A_{b}^{z}(q)({\bf A}_{c}^{\perp}(q')\times{\bf A}_{a}^{\perp}(p))^{z}\nonumber \\
 & \times(G_{\sigma}^{R}(z+\omega)G_{\sigma}^{R}(z)G_{\bar{\sigma}}^{A}(z)+G_{\sigma}^{R}(z)G_{\sigma}^{A}(z)G_{\bar{\sigma}}^{A}(z-\omega)+(b\leftrightarrow c)\nonumber \\
= & \sum_{\sigma}\frac{e^{3}}{16\pi m^{3}}A_{b}^{z}(q')F_{ac}(p,q)\int\nu(\varepsilon)d\varepsilon(G_{\sigma}^{R}(\omega)G_{\sigma}^{R}G_{\bar{\sigma}}^{A}+G_{\sigma}^{R}G_{\sigma}^{A}G_{\bar{\sigma}}^{A}(-\omega)+(b\leftrightarrow c)
\end{align}
and 
\begin{align}
 & \mathcal{D}_{abc}^{2(f)}(Q',\omega)\nonumber \\
= & \sum_{\sigma}\frac{ie^{3}\sigma}{16\pi m^{3}}\int dzn'_{F}(z)i\sigma\int[dk]A_{b}^{z}(q)({\bf A}_{a}^{\perp}(p)\times{\bf A}_{c}^{\perp}(q'))^{z}\nonumber \\
 & \times(G_{\bar{\sigma}}^{R}(z+\omega)G_{\sigma}^{R}(z)G_{\sigma}^{A}(z)+G_{\bar{\sigma}}^{R}(z)G_{\sigma}^{A}(z)G_{\bar{\sigma}}^{A}(z-\omega))+(b\leftrightarrow c)\nonumber \\
= & -\sum_{\sigma}\frac{e^{3}}{16\pi m^{3}}A_{b}^{z}(q')F_{ac}(p,q)\int\nu(\varepsilon)d\varepsilon(G_{\bar{\sigma}}^{R}(\omega)G_{\sigma}^{R}G_{\sigma}^{A}+G_{\bar{\sigma}}^{R}G_{\sigma}^{A}G_{\sigma}^{A}(-\omega)+(b\leftrightarrow c)
\end{align}
in which $Q'=p+q+q'$ and we ingore the index $k$ in the Green's
function here. Then we have 
\begin{align}
 & \Xi{}_{abc}^{2(d)+(f)}(Q')\nonumber \\
\approx & \frac{\partial}{\partial\omega}\sum\frac{ie^{3}\sigma}{16\pi m^{3}}\int dzn'_{F}(z)\int[dk]i\sigma\nonumber \\
 & \times[A_{b}^{z}(q)({\bf A}_{a}^{\perp}(q')\times{\bf A}_{c}^{\perp}(p))^{z}(G_{\sigma}^{R}(z+\omega)G_{\sigma}^{R}(z)G_{\bar{\sigma}}^{A}(z)+G_{\sigma}^{R}(z)G_{\sigma}^{A}(z)G_{\bar{\sigma}}^{A}(z-w)\nonumber \\
 & -A_{b}^{z}(q)({\bf A}_{a}^{\perp}(p)\times{\bf A}_{c}^{\perp}(q'))^{z}(G_{\bar{\sigma}}^{R}(z+\omega)G_{\sigma}^{R}(z+\omega)G_{\sigma}^{A}(z)+G_{\bar{\sigma}}^{R}(z)G_{\sigma}^{A}(z)G_{\sigma}^{A}(z-\omega)]|_{\omega=0}\nonumber \\
 & +(b\leftrightarrow c)\nonumber \\
= & \sum_{\sigma}\frac{1}{16\pi}(\frac{e}{m})^{3}\int[dk]K(F_{ac}(p,q')A_{b}^{z}(q)+F_{ab}(p,q')A{}_{c}^{z}(q)]+(b\leftrightarrow c)
\end{align}
in which $K=\int[dk](G_{\sigma}^{R3}G_{\bar{\sigma}}^{A}-G_{\bar{\sigma}}^{R2}G_{\sigma}^{R}G_{\sigma}^{A}-G_{\sigma}^{R}G_{\sigma}^{A}G_{\bar{\sigma}}^{A2}+G_{\bar{\sigma}}^{R}G_{\sigma}^{A3})$.
Then we first calculate
\begin{align}
K_{1} & =\int_{0}^{\infty}\frac{\nu(\varepsilon)d\varepsilon}{(\varepsilon-\mu_{\sigma}-i\eta)^{3}(\varepsilon-\mu_{\bar{\sigma}}+i\eta)}\nonumber \\
 & \approx C\pi i[\frac{\sqrt{\mu_{\bar{\sigma}}-i\eta}}{(\mu_{\bar{\sigma}}-\mu_{\sigma}-2i\eta)^{3}}+\frac{\sqrt{\mu_{\sigma}+i\eta}}{(\mu_{\sigma}-\mu_{\bar{\sigma}}+2i\eta){}^{3}}-\frac{1}{2(2\sigma M+2i\eta)^{2}\sqrt{\mu_{\sigma}}}]\nonumber \\
 & \approx\frac{\pi i}{8}\frac{\nu_{\sigma}-\nu_{\bar{\sigma}}}{(\sigma M+i\eta)^{3}}.
\end{align}
Then we calculate 
\begin{align}
K_{4} & =\int_{0}^{\infty}\frac{\nu(\varepsilon)d\varepsilon}{(\varepsilon-\mu_{\bar{\sigma}}-i\eta)(\varepsilon-\mu_{\sigma}+i\eta)^{3}}\nonumber \\
 & \approx C\pi i[\frac{\sqrt{\mu_{\bar{\sigma}}+i\eta}}{(\mu_{\bar{\sigma}}-\mu_{\sigma}+2i\eta)^{3}}+\frac{\sqrt{\mu_{\sigma}-i\eta}}{(\mu_{\sigma}-\mu_{\bar{\sigma}}-2i\eta)^{3}}-\frac{1}{2(\mu_{\sigma}-\mu_{\bar{\sigma}}-2i\eta)^{2}\sqrt{\mu}}]\nonumber \\
 & \approx\frac{\pi i}{8}\frac{\nu_{\sigma}-\nu_{\bar{\sigma}}}{(\sigma M-i\eta)^{3}}.
\end{align}
And we get 
\begin{equation}
K_{1}+K_{4}\approx\frac{\pi i\sigma M(\nu_{\sigma}-\nu_{\bar{\sigma}})}{4}\frac{(M^{2}-3\eta^{2})}{(M^{2}+\eta^{2})^{3}}.
\end{equation}
Next is 
\begin{align}
K' & =K_{2}+K_{3}\nonumber \\
 & =-\frac{C}{2i\eta}\int_{0}^{\infty}\sqrt{\varepsilon}d\varepsilon(\frac{1}{\varepsilon-\mu_{\sigma}-i\eta}-\frac{1}{\varepsilon-\mu_{\sigma}+i\eta})[\frac{1}{(\varepsilon-\mu_{\bar{\sigma}}-i\eta)^{2}}+\frac{1}{(\varepsilon-\mu_{\bar{\sigma}}+i\eta)^{2}}]\nonumber \\
 & \approx\frac{2\pi iM\sigma}{(M^{2}+\eta^{2})^{2}}(\nu_{\sigma}-\nu_{\bar{\sigma}}).
\end{align}
After summation over $\sigma$, the contribution is zero at linear
$\nu_{\sigma}$ order. 

\bibliographystyle{unsrt}
\bibliography{nlhc}